\documentclass[aps,prc,12pt,amsmath,amssymb,nofootinbib,superscriptaddress,preprint]{revtex4-1}
\usepackage{selinput}
\usepackage[T1]{fontenc}
\usepackage{graphicx}
\usepackage{mathtools}
\usepackage{verbatim}
\usepackage[dvipsnames]{xcolor}
\usepackage{geometry}
\usepackage{siunitx}
\usepackage[toc,page]{appendix}\usepackage{float}
\usepackage{subcaption}
\usepackage{lipsum}
\usepackage[breaklinks, colorlinks=true, citecolor=blue]{hyperref}
\usepackage{caption}
\usepackage{bm}

\begin{document}
\title{Has Quarkyonic Matter Been Seen in Lattice Simulations?}

\author{Kohei~Fujikura}
\affiliation{Yukawa Institute for Theoretical Physics, Kyoto University, Kyoto 606-8502, Japan}

\author{Kenji~Fukushima}
\affiliation{Department of Physics, The University of Tokyo, 7-3-1 Hongo, Bunkyo-ku, Tokyo 113-0033}

\author{Yoshimasa~Hidaka}
\affiliation{Yukawa Institute for Theoretical Physics, Kyoto University, Kyoto 606-8502, Japan}
\affiliation{RIKEN Center for Interdisciplinary Theoretical and Mathematical Sciences (iTHEMS), RIKEN, Wako 351-0198, Japan}

\author{Etsuko~Itou}

\affiliation{Yukawa Institute for Theoretical Physics, Kyoto University, Kyoto 606-8502, Japan}

\author{Larry~McLerran}
\affiliation{Institute for Nuclear Theory, University of Washington, Box 351550, Seattle, WA 98195, USA}

\preprint{YITP-26-153, RIKEN-iTHEMS-Report-26}

\newcommand{\LM}[1]{\textcolor{Red}{#1}}
\newcommand{\EI}[1]{{\color{red}{#1}}}
\newcommand{\kf}[1]{{\color{magenta}#1}}
\newcommand{\YH}[1]{{\color{blue}{#1}}}
\definecolor{amethyst}{HTML}{9966CC}
\newcommand{\fujiQ}[1]{{\color{amethyst}{#1}}}

\newcommand{\LQCD}{\Lambda_\mathrm{QCD}}
\newcommand{\Tc}{T_{\mathrm{c}}}
\newcommand{\Td}{T_{\mathrm{d}}}
\newcommand{\Th}{T_{\mathrm{H}}}
\newcommand{\ri}{\mathrm{i}}
\newcommand{\Nc}{N_\mathrm{c}}
\newcommand{\Nf}{N_\mathrm{f}}
\newcommand{\Nl}{N_{\mathrm{l}}}
\newcommand{\Tr}{\mathop{\mathrm{Tr}}}
\renewcommand{\Re}{\mathop{\mathrm{Re}}}
\newcommand{\muQ}{\mu_\mathrm{Q}}
\newcommand{\muB}{\mu_\mathrm{B}}
\newcommand{\muI}{\mu_\mathrm{I}}
\newcommand{\nB}{n_\mathrm{B}}
\newcommand{\nI}{n_\mathrm{I}}
\newcommand{\muBhat}{\hat{\mu}_\mathrm{B}}
\newcommand{\MB}{M_\mathrm{B}}
\newcommand{\qbul}{q_{\mathrm{bulk}}}
\newcommand{\llimit}{\nearrow}
\newcommand{\bk}{{\bm k}}
\newcommand{\bp}{{\bm p}}
\newcommand{\bq}{{\bm q}}
\newcommand{\Kone}{K_{\text{1+1}}}

\begin{abstract}
  Quarkyonic Matter refers to the dual state of matter between confined hadrons and quasi-free quarks.  This duality leads to a robust constraint on the behavior of the distribution functions.  After a comprehensive review of Quarkyonic Matter based on the distribution functions, we discuss recent lattice data for two-color QCD in $(1+1)$ and $(3+1)$ dimensions at finite baryon number density and QCD in $(3+1)$ dimensions at finite isospin density, as well as the astrophysically inferred equation of state.  
  Taken together, these results reveal a coherent picture consistent with the emergence of Quarkyonic Matter in lattice simulations. Future lattice calculations of baryon and gauge-invariant quark momentum distributions at high density would provide a more direct test of this interpretation.
\end{abstract}

\maketitle

\section{Introduction}
\label{sec:intro}

Quantum Chromodynamics (QCD) is the fundamental theory of the strong interaction with elementary degrees of freedom of quarks and gluons.  Phases of hot QCD matter can be accessed by first-principles lattice QCD simulations, whereas finite-density simulations are problematic due to the sign problem~\cite{Nagata:2021ugx, Aarts:2026uiu}.  Although there are various theoretical speculations about the realizations of exotic phases of dense QCD matter, no one would doubt the existence of quark matter in the limit of asymptotically high density~\cite{Itoh:1970uw, Collins:1974ky}; possibly it may even exist in the deepest cores of massive neutron stars~\cite{Annala:2019puf}.

The theoretically challenging question is what quark matter should be at all.  In fact, no symmetry-based argument works to characterize whether quarks are confined or deconfined with dynamical quark degrees of freedom.  Even though no clear definition of quark matter has been established, we can still discuss thermodynamic properties of matter.  It would be a reasonable working hypothesis that quark matter is realized when thermodynamic properties are well approximated by a gas of weakly interacting quarks.  This characterization is not founded on a rigorous quantum field theory argument, but it should be phenomenologically acceptable.

Let us take a further step toward the microscopic dynamics in the intermediate stage from confined hadronic matter to deconfined quark matter.  In the simplest interpolation prescription, two regimes are connected with a smooth curve~\cite{Masuda:2012ed, Masuda:2012kf} (see also Ref.~\cite{Alvarez-Castillo:2013spa} with some interpretations), unless a first-order phase transition causes discontinuous behavior in thermodynamics.  If one sees the QCD phase diagram research, one can find various microscopic mechanisms to favor a first-order phase transition~\cite{Fukushima:2010bq, Baym:2017whm}, but for the moment there is no theoretical nor experimental evidence to support the existence of the first-order phase transition.  The QCD Critical Point search is still ongoing with help of lattice QCD simulations (see, e.g., Ref.~\cite{Borsanyi:2025dyp} for the constant entropy density, Ref.~\cite{Clarke:2024ugt} for the Lee-Yang edge singularities, etc.) and the functional QCD methods (see, e.g., Ref.~\cite{Fischer:2026uni} for a review, Ref.~\cite{Fu:2026qnl} for the latest estimate, etc.).

Some additionally constraining information is experimentally available from neutron-star measurements, especially the mass-radius distribution.  The discovery of heavy neutron stars as massive as two solar mass~\cite{Demorest:2010bx, Antoniadis:2013pzd} has been the turning point; since then, the studies of neutron-star matter equation-of-state (EoS), that is the relation between pressure $p$ and energy density $\varepsilon$, have significantly developed~\cite{Alford:2015dpa, Baym:2017whm, Annala:2017llu, Annala:2019puf}.  The crucial message from those studies is that the EoS must be stiff enough to support heavy neutron stars, and accordingly the speed of sound, $v_s^2=dp/d\varepsilon$, must become large~\cite{Kojo:2014rca, Bedaque:2014sqa, Tews:2018kmu}; in the crossover EoS picture, the enhancement of $v_s$ was pointed out already in Ref.~\cite{Masuda:2012ed} and emphasized recently in Ref.~\cite{Fujimoto:2022xhv}.  This observation brings about a surprising consequence.  Since the EoS of nuclear matter turns out to be soft according to chiral effective theory or chiral perturbation theory (ChPT)~\cite{Fiorilla:2011sr, Hebeler:2013nza, Drischler:2020fvz}, the EoS should be such  that $v_s^2$ can exceed the conformal value of $1/3$ at intermediate densities realized inside the neutron stars.  At asymptotically high density, QCD approaches conformal behavior~\cite{Freedman:1976ub, Kurkela:2009gj, Gorda:2021znl} (it should be noted that the neutron-star matter quickly exhibits conformality; see~Ref.~\cite{Fujimoto:2022ohj}), and thus, $v_s^2\to 1/3$ holds eventually in the high-density limit.  Therefore, the neutron-star constraint implies that $v_s^2$ exhibits a peak above the conformal value at some density which is around several times the normal nuclear density~\cite{Fujimoto:2019hxv, Altiparmak:2022bke, Brandes:2022nxa, Marczenko:2022jhl, Fujimoto:2024cyv}.

\begin{figure}
    \centering
    \includegraphics[width=0.9\textwidth]{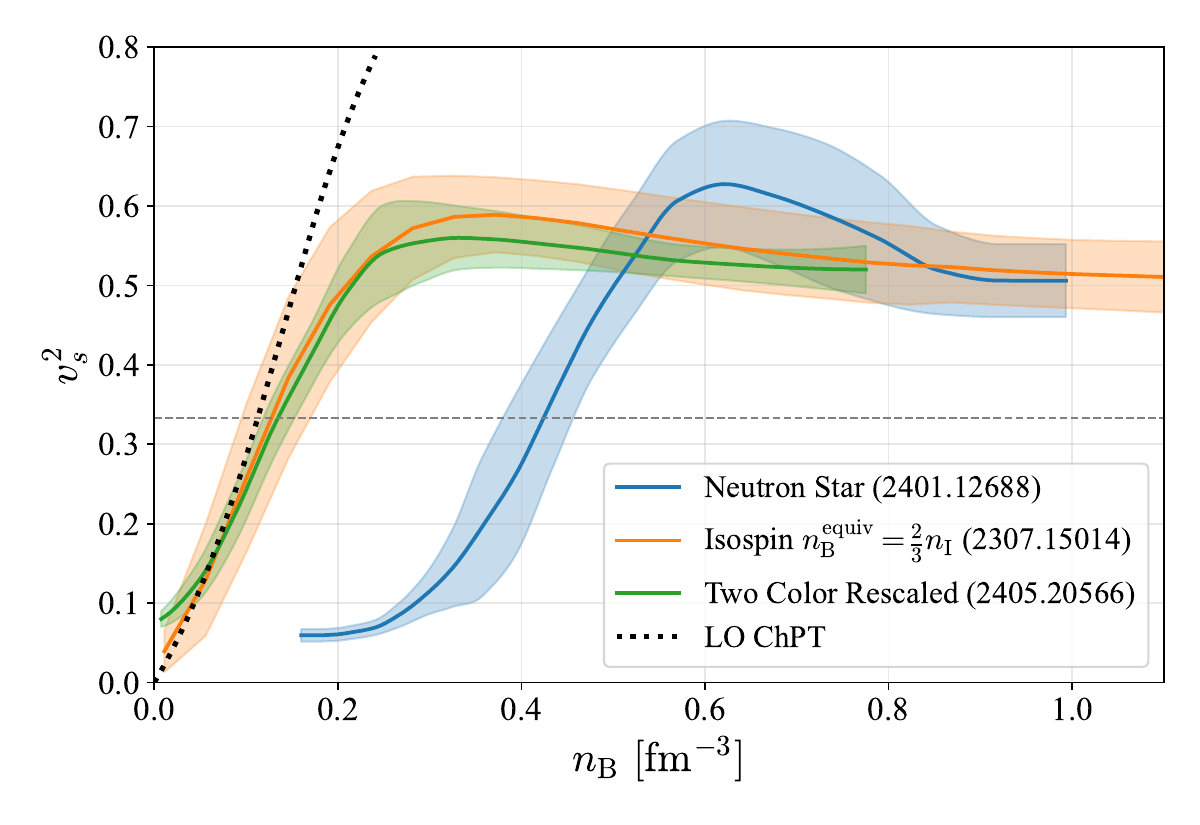}
    \caption{Speed of sound obtained in various physical systems.  The blue band represents the data inferred from the neutron-star measurements~\cite{Fujimoto:2024cyv}, while the green and the orange bands represent the lattice simulation results at high baryon density of two-color QCD~\cite{Iida:2024irv} and high isospin density of three-color QCD~\cite{Abbott:2023coj}, respectively.  The dotted curve is the leading-order ChPT prediction, Eq.~\eqref{eq:ChPT_vs}, and the dashed line indicates the conformal value, $v_s^2 = 1/3$.  The rescaling factors for comparison are explained in the text; see Eqs.~\eqref{eq:nB_equiv} and \eqref{eq:nB_equiv_2c}.}
    \label{fig:vs_comparison}
\end{figure}

The peaked structure of $v_s^2$ with increasing baryon density, $\nB$, as plotted by the blue band in Fig.~\ref{fig:vs_comparison}, poses a critical challenge in high-density QCD research.  Typically, phase-transitional behavior would not enhance but suppress $v_s^2$;  indeed, $v_s^2=0$ for the Maxwell-constructed EoS with a first-order phase transition.  From the experimental data, we inevitably conclude that something more nontrivial than a conventional phase-transition picture occurs there.  In the traditional many-body calculation of nuclear matter, rapid stiffening cannot be derived, and we need to understand the precursory physics at intermediate density from baryonic to quark matter, which lies in the density regime of Quarkyonic Matter~\cite{McLerran:2007qj, McLerran:2018hbz}.

In this paper, we will give a comprehensive overview of the mechanism to account for the sound speed peak based on the Quarkyonic Matter picture~\cite{McLerran:2018hbz, Jeong:2019lhv, Duarte:2020xsp, Zhao:2020dvu, Cao:2020byn, Margueron:2021dtx}.  Importantly, the onset of Quarkyonic Matter is recently defined by a quarkyonic saturation condition~\cite{Kojo:2021ugu, Fujimoto:2023mzy}, which leads to an unambiguous picture of quarkyonic duality.  Therefore, Quarkyonic Matter is not merely a conceptual characterization, but it is a well-defined object.

Inspired by the astrophysically constrained $v_s^2$, high-density QCD-like theories free from the sign problem have been investigated on the lattice~\cite{Hands:2006ve, Cotter:2012mb, Braguta:2016cpw, Astrakhantsev:2018uzd, Boz:2019enj, Iida:2019rah, Begun:2022bxj, Brandt:2017oyy, Brandt:2022hwy, Abbott:2023coj}.  Indeed, in these setups, it has been numerically confirmed that the speed of sound can exceed the conformal value at rather low densities~\cite{Iida:2022hyy, Iida:2024irv, Brandt:2022hwy, Abbott:2023coj}.  More specifically, in the present work, we will discuss the lattice results for two-color QCD at high baryon density and three-color QCD at high isospin density.  However, we should be cautious about the physical interpretation.  The dense two-color and high-isospin-density systems are dominated by diquarks and pions, respectively, allowing for a model-independent ChPT description of low-energy properties.  In fact, the increasing trend of $v_s^2$ is derived from ChPT as we explain below.

From leading-order ChPT, the theoretical expressions for the isospin system are given as~\cite{Son:2000xc, Son:2000by}
\begin{equation}
  v_s^2 = \frac{\nI}{\muI d\nI/d\muI} \,, 
  \qquad
  \nI = \frac{f_\pi^2}{2} \muI \biggl( 1 - \frac{m_\pi^4}{\muI^4} \biggr)
\end{equation}
for $\muI \ge m_\pi$.\footnote{ We normalize $f_\pi$ as in Ref.~\cite{Abbott:2023coj} where the physical value is $f_\pi^{\text{(phys)}}\approx \qty{130}{\MeV}$.  This $f_\pi$ is $\sqrt{2}$ times the older convention of $F_\pi\approx \qty{93}{\MeV}$ as in Ref.~\cite{Son:2000xc, Son:2000by}.}  We introduce a parametric variable, $x=\muI/m_\pi$, and then the speed of sound is rewritten as
\begin{equation}
    v_s^2(x) = \frac{1-x^{-4}}{1+3x^{-4}}\,.
    \label{eq:ChPT_vs}
\end{equation}
To compare with the data in terms of baryon density, we count the quarks and antiquarks in the condensed pions.  The $\pi^+$ consists of a $u$ quark and a $\bar{d}$ antiquark, so that $n_u + n_{\bar{d}} = 2\nI$. Dividing the total quark number, $n_u + n_{\bar{d}}$, by $\Nc=3$, we define the ``equivalent baryon density'' as
\begin{equation}
    n_{\mathrm{B}}^{\text{equiv}}(x) = \frac{2\nI(x)}{3} = \frac{f_\pi^2 m_\pi}{3} x (1-x^{-4})\,.
    \label{eq:nB_equiv}
\end{equation}
Therefore, from these expressions, we can draw the theoretical curve $(n_{\mathrm{B}}^{\text{equiv}}(x),\, v_s^2(x))$ by varying $x$.  In Fig.~\ref{fig:vs_comparison}, the high-isospin lattice data are plotted as a function of $\nB^{\text{equiv}}=\tfrac{2}{3}\nI$.

The two-color case is parallel to the isospin case.  In two-color QCD, the lightest baryon is a diquark, which belongs to the same chiral multiplet as the pion~\cite{Kogut:2000ek, Hands:2006ve}.  Condensed diquarks play the role of the condensed $\pi^+$.  Accordingly, $\muB$, $m_{\mathrm{PS}}$, and $f_{\mathrm{PS}}$ replace $\muI$, $m_\pi$, and $f_\pi$, where $m_{\mathrm{PS}}$ and $f_{\mathrm{PS}}$ are the mass and the decay constant of the lightest pseudoscalar.\footnote{In our convention, $f_{\mathrm{PS}}=2\sqrt{2}F$ where $F$ is the two-color counterpart of $F_\pi$ used in Refs.~\cite{Kogut:2000ek, Hands:2006ve, Iida:2024irv}.}  Thus, with $x'=\muB/m_{\mathrm{PS}}$, $v_s^2(x')$ takes the same form as Eq.~\eqref{eq:ChPT_vs}~\cite{Hands:2006ve}, and the baryon number $n_{\mathrm{B}}^{\text{(2c)}}(x')$ plays the role of $\nI(x)$.  The only difference is the scale, $f_{\mathrm{PS}}^2 m_{\mathrm{PS}}$ instead of $f_\pi^2 m_\pi$.  As we did previously, we divide the total quark number, $2\nB^{\text{(2c)}}$ by $\Nc=3$, to find the equivalent baryon density as in Eq.~\eqref{eq:nB_equiv}, i.e.,
\begin{equation}
    \nB^{\text{equiv}}(x') = \frac{2}{3}\, \alpha\, \nB^{\text{(2c)}}(x')\,, \qquad
    \alpha = \frac{f_\pi^2 m_\pi}{f_{\mathrm{PS}}^2 m_{\mathrm{PS}}}\,.
    \label{eq:nB_equiv_2c}
\end{equation}
In the lattice-QCD simulation at high isospin density~\cite{Abbott:2023coj}, these parameters are close to the physical values: $f_\pi=\qty{132}{\MeV}$ and $m_\pi=\qty{170}{\MeV}$.  In the two-color case~\cite{Iida:2024irv} with a conversion factor, $0.75$, corresponding to the choice of $\Tc=\qty{150}{\MeV}$ (see Sec.~\ref{sec:2c}), $f_{\mathrm{PS}}=0.75\times2\sqrt{2}\times\qty{56.7}{\MeV}\approx\qty{120}{\MeV}$ and $m_{\mathrm{PS}}=0.75\times\qty{738}{\MeV} \approx \qty{553.5}{\MeV}$.  Then, we find $\alpha\approx 0.370$.

Figure~\ref{fig:vs_comparison} shows that leading-order ChPT agrees well with both the isospin and two-color (rescaled by $2\alpha/3 \approx 0.247$) lattice data up to $\nB \lesssim \qty{0.1}{\femto\meter^{-3}}$.  Unlike the sound speed peak in nuclear/neutron matter, ChPT explains why the speed of sound increases rapidly and exceeds the conformal value.  Therefore, rather than the increasing behavior, the most nontrivial feature is how the speed of sound deviates from the ChPT prediction at high density.  Interestingly, even after the deviation from the ChPT prediction for $\nB\gtrsim \qty{0.1}{\femto\meter^{-3}}$, the isospin and rescaled two-color lattice data remain close to each other.  This observation suggests that the underlying mechanism of the sound speed should be common for these distinct physical systems.
We argue that the underlying mechanism could be a realization of Quarkyonic Matter.

\section{Physics Backgrounds}
\label{sec:background}

It is a ubiquitous phenomenon that the low-energy excitations in a many-body system are different from those of the elementary particles of the theory.  For example, in a superconductor, the condensate consists of bosonic Cooper pairs, while the constituent particles are fermionic electrons.  In a Fermi liquid, the low-energy excitations are quasi-particles with properties similar to those of the constituent fermions, while in one spatial dimension this quasi-particle description breaks down for any nonzero interaction.  A typical example is the Tomonaga--Luttinger liquid.  At low energies, in this case, the excitations are bosonic collective particle-hole modes.  Interestingly, the Tomonaga--Luttinger liquid does not lose information about the constituent fermions~\cite{Voit:1995}.  The Fermi momentum is still fixed by the density as for free fermions, in agreement with Luttinger's theorem, although the momentum distribution has only a power-law singularity there instead of a jump.  The Tomonaga--Luttinger liquid has been observed in a number of condensed matter systems, e.g., in carbon nanotubes~\cite{Bockrath:1999} and through spin-charge separation in semiconductor quantum wires~\cite{Auslaender:2005}.  In QCD in $(1+1)$ dimensions, in contrast, the color degrees of freedom are gapped, and the low-energy sector is a Tomonaga--Luttinger liquid of the color-singlet baryon density~\cite{Lajer:2021kcz}.  Nevertheless, the quark distribution develops a Fermi sea at high density, as we will see in Sec.~\ref{sec:lattice}.  This coexistence of a quark Fermi sea with color-singlet excitations is the dual character of Quarkyonic Matter.  In this way, the idea of Quarkyonic Matter is linked to various physical systems, and it can be realized in QCD-like theories.

\subsection{Quarkyonic Matter}

Quarkyonic Matter was proposed to resolve a paradox arising for QCD at finite baryon number density in the large-$\Nc$ limit~\cite{McLerran:2007qj}.  Let us explain the problem.  The leading-order Debye screening mass is
\begin{equation}
    M^2_\mathrm{Debye} = \frac{g^2}{2\pi^2} \sum_{f=1}^{\Nf} \mu_f^2 \sim \frac{\lambda\, \Nf}{\Nc} \, \muQ^2\,,
    \label{eq:Debye_QCD}
\end{equation}
where $\muQ$ is the quark chemical potential, $\Nc$ is the number of colors, $g$ is the QCD coupling constant, and $\lambda = g^2 \Nc$ is the 't~Hooft coupling which remains of $\mathcal{O}(1)$ in the $\Nc\to \infty$ limit.  For a free Fermi gas, $\muQ$ is nothing but the Fermi energy of quarks.  Confinement persists as long as the screening length is longer than the confinement scale, $1/\LQCD$.  As we see from Eq.~\eqref{eq:Debye_QCD}, $M_\mathrm{Debye} \ll \LQCD$ as long as $\muQ \ll \sqrt{\Nc}\,\LQCD$.  Thus, the theory remains confined unless $\muQ$ becomes as large as $\sqrt{\Nc}\,\LQCD$.  As a consequence of asymptotic freedom in QCD, the interaction strength of quarks should be weak when $\muQ \gg \LQCD$.  Therefore, for $\LQCD \ll \muQ \ll \sqrt{\Nc}\,\LQCD$ in the large-$\Nc$ limit, we would na\"{i}vely expect matter to be a weakly interacting gas of quarks, in contradiction to the argument based on Eq.~\eqref{eq:Debye_QCD}.  

\begin{figure}
    \centering
    \includegraphics[width=0.3\textwidth]{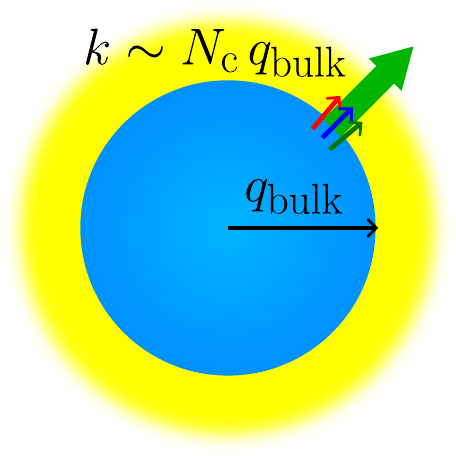}
    \caption{Schematic picture of Quarkyonic Matter.  The bulk property of the Fermi sea is dominated up to $q\sim \qbul$ by degenerate quarks, while excitations on the Fermi sea are confined baryons that carry the momentum, $k\sim \Nc \qbul$.}
    \label{fig:quarkyonic}
\end{figure}

The paradox is resolved if dense matter has different degrees of freedom at the Fermi surface and deep inside the Fermi sea.
The exchange interactions deep inside the Fermi sea are not very sensitive to confining infrared scales because of Pauli blocking.  These quarks are quasi-free due to weak interactions at $\muQ \gg \LQCD$, although matter as a whole remains confined.  In contrast, interaction effects at the Fermi surface are sensitive to processes with soft-momentum exchange.  In the QCD case, such processes at the Fermi surface bind quarks into baryons for $\muQ \ll \sqrt{\Nc}\,\LQCD$.  Accordingly, quark particle-hole excitations should form confined mesons, and gluons should be confined into glueballs.  Such matter was named Quarkyonic after its properties as both quark matter, whose pressure is of order $\Nc$ like that of quarks, and baryonic matter, whose excitations near the Fermi surface are confined baryons~\cite{McLerran:2007qj}.  Schematically, at low temperatures, Quarkyonic Matter has a momentum shell structure of the quark Fermi sea and the surrounding baryon excitations as illustrated in Fig.~\ref{fig:quarkyonic}.

It is important to note that Quarkyonic Matter has a phenomenological impact particularly in understanding neutron star structures.  The description based on Quarkyonic Matter at low temperature was successful in generating a stiff EoS needed to explain the observation~\cite{McLerran:2018hbz}.  Essential in this Quarkyonic Matter description is a conversion mechanism associated with a crossover transition between nonrelativistic baryonic matter and relativistic quarks in Quarkyonic Matter in a small range of density.

\subsection{Distribution Functions in a Quarkyonic Picture}

A specific model for Quarkyonic Matter was proposed in Ref.~\cite{Fujimoto:2023mzy}.  It is assumed that nucleons and quarks are treated as noninteracting particles except for confinement. In this model, the confined nucleon is simplified as the bound state of quarks with equal momenta. Thus, the phase-space density or the momentum-space distribution function of quarks and that of nucleons are related by
\begin{equation}
    f_\mathrm{Q}(\bq) = \int \frac{d^3k}{(2\pi)^3}\, K_N(\bq-\bk/\Nc) f_\mathrm{B}(\bk) \,,
    \label{eq:idylliq}
\end{equation}
where $K_N(q)$ is the probability density for finding a quark inside a nucleon at rest.  Above the saturation density discussed below, $f_{\mathrm{B}}$ is determined from $f_{\mathrm{Q}}$ through Eq.~\eqref{eq:idylliq} with a fixed $K_N$.  It coincides with the momentum distribution of nucleons only when the nucleon wave function is unchanged by the medium.  The quantity directly compared with lattice data in Sec.~\ref{sec:lattice} is $f_{\mathrm{Q}}$.  We note that, throughout this paper, $\bq$ denotes the momentum of quarks, while $\bk$ represents the baryon momentum.
It is a natural assumption that $K_N(\bq)$ is peaked around $|\bq|=0$ with a finite width parametrized by $\Lambda$.
Under the normalization condition,
\begin{equation}
    \int {{d^3q} \over {(2\pi)^3}} K_N(\bq) = 1\,,
    \label{eq:KN_norm}
\end{equation}
the simplest choice is the exponential form~\cite{Fujimoto:2023mzy}:
$K_N(\bq) = (2\pi^2/\Lambda^3) e^{-|\tilde{\bq}|}/|\tilde{\bq}|$ with $\tilde{\bq}=\bq/\Lambda$.  It should be noted that, in this normalization convention, the integration with respect to $\bq$ leads to
\begin{equation}
    \int \frac{d^3 q}{(2\pi)^3} \, f_{\mathrm{Q}}(\bq)
    = \int \frac{d^3 k}{(2\pi)^3} \, f_{\mathrm{B}}(\bk)\,.
    \label{eq:fQB}
\end{equation}
We should remember that $f_{\mathrm{Q}}$ and $f_{\mathrm{B}}$ represent the distribution functions of quarks and baryons, respectively, per degree of freedom.  Thus, the corresponding number densities with $\Nc$ colors, $\Nf=2$ flavors, and a factor of 2 from spin degeneracy are
\begin{equation}
    n_{\mathrm{Q}} = 4\Nc \int \frac{d^3 q}{(2\pi)^3}\, f_{\mathrm{Q}}(\bq)\,,
    \qquad
    n_{\mathrm{B}} = 4 \int \frac{d^3 k}{(2\pi)^3}\, f_{\mathrm{B}}(\bk)\,,
    \label{eq:nQB}
\end{equation}
with which Eq.~\eqref{eq:fQB} is consistent with $n_{\mathrm{Q}}=\Nc\,n_{\mathrm{B}}$.

\begin{figure}
    \centering
    \includegraphics[width=0.80\textwidth]{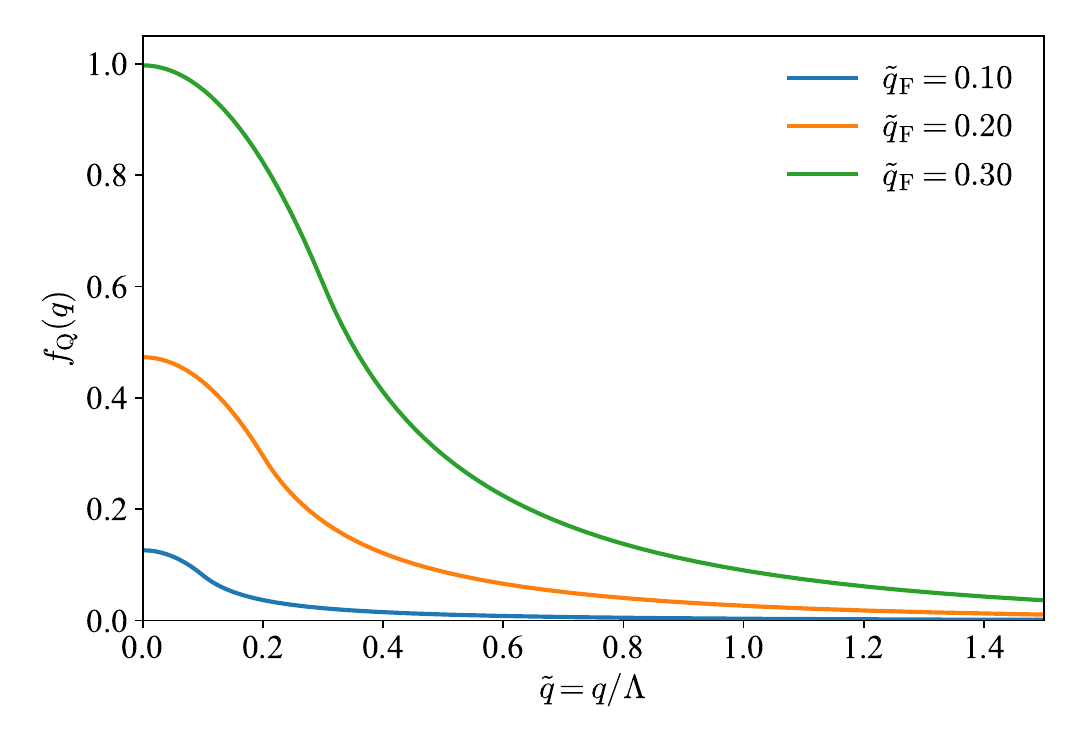}
    \caption{$f_{\mathrm{Q}}(q)$ for $\Nc=3$ given in Eq.~\eqref{eq:fQana} with increasing $\tilde{q}_{\mathrm{F}}$ before quarkyonic saturation.}
    \label{fig:fQ}
\end{figure}

Since quarks and baryons are fermions, their maximum occupation number per phase-space cell and degree of freedom is one.  For simplicity, we take a Fermi--Dirac distribution for the nucleons at $T=0$, i.e.,
\begin{equation}
  f_\mathrm{B}(\bk) = \Theta(\muB - E_k)\,,
  \label{eq:fB_Theta}
\end{equation}
where $\Theta(x)$ is the Heaviside step function and $E_k = \sqrt{\bk^2 + m_N^2}$ with the nucleon mass $m_N$.  From this, we define the Fermi momentum of nucleons, that is, $k_{\mathrm{F}} = \sqrt{\muB^2 - m_N^2}$ for $\muB \ge m_N$.  Then, Eq.~\eqref{eq:idylliq} is analytically integrated, leading to
\begin{equation}
  f_{\mathrm{Q}}(q) = \Nc^3
  \begin{cases}
  \displaystyle
      1 - (\tilde{q}_{\mathrm{F}}+1) e^{-\tilde{q}_{\mathrm{F}}} \frac{\sinh \tilde{q}}{\tilde{q}} & \tilde{q} \leq \tilde{q}_{\mathrm{F}}\,, \\
  \displaystyle
      \frac{e^{-\tilde{q}}}{\tilde{q}}(\tilde{q}_{\mathrm{F}} \cosh \tilde{q}_{\mathrm{F}} - \sinh \tilde{q}_{\mathrm{F}}) & \tilde{q} > \tilde{q}_{\mathrm{F}}\,,
  \end{cases}
  \label{eq:fQana}
\end{equation}
with $\tilde{q}=q/\Lambda$ and $\tilde{q}_{\mathrm{F}} = k_{\mathrm{F}}/(\Nc \Lambda) = q_{\mathrm{F}}/\Lambda$, where $q_{\mathrm{F}} = k_{\mathrm{F}}/\Nc$ is the quark momentum corresponding to the nucleon Fermi momentum.  As plotted in Fig.~\ref{fig:fQ}, the quark distribution function rises as the Fermi momentum increases.

There will be some critical Fermi momentum, $q_{\mathrm{F}}^{\text{sat}}$, at which $f_{\mathrm{Q}}(0)$ equals one.  For $q_{\mathrm{F}} > q_{\mathrm{F}}^{\text{sat}}$, the assumed form of the nucleon distribution function in Eq.~\eqref{eq:fB_Theta} is no longer valid.  This criterion of ``quarkyonic saturation'' is called the Kojo Filling Criterion (KFC)~\cite{Kojo:2021ugu}, which can be written down from Eq.~\eqref{eq:idylliq} as
\begin{equation}
   1 = \int~{{d^3 k} \over {(2\pi)^3}} K_N(\bk/\Nc) f_{\mathrm{B}}(\bk)\,.
\end{equation}
In the case of the exponential kernel, the KFC takes a simple form of $1=\Nc^3[1-(1+x)e^{-x}]$ with $x=k_{\mathrm{F}}/(\Nc\Lambda)$.  Therefore, for $\Nc\gg 1$, the solution reads: $k_{\mathrm{F}} \simeq \sqrt{2/\Nc}\,\Lambda$~\cite{Fujimoto:2023mzy} and the corresponding density is
\begin{equation}
    \nB^{\text{sat}} \simeq \frac{2^{5/2}}{3\pi^2} \Lambda^3 \Nc^{-3/2} \,.
\end{equation}
If we choose $K_N(\bq)$ such that it is not singular at $|\bq|=0$, taking the large-$\Nc$ limit reduces the KFC to
\begin{equation}
   1 = K_N(0)\, \frac{n_{\mathrm{B}}^{\text{sat}}}{4} \,.
   \label{eq:KFC_QCD}
\end{equation}   
In reasonable models of the quark substructure of the nucleon, $n_{\mathrm{B}}^{\text{sat}} = n_{\mathrm{Q}}(q_{\mathrm{F}}^{\text{sat}})/\Nc$ is estimated to be close to the saturation density of normal nuclear matter~\cite{Koch:2024qnz,McLerran:2024rvk}.

\begin{figure}
    \centering
    \includegraphics[width=0.9\textwidth]{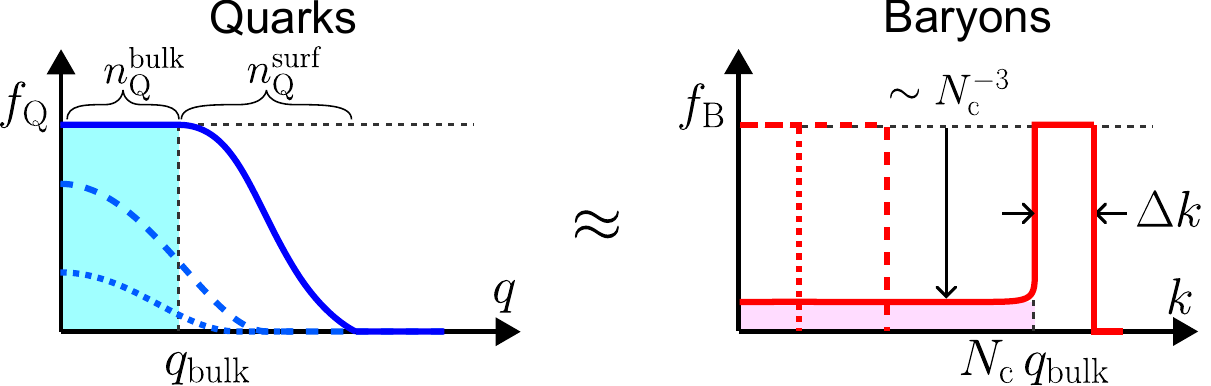}
    \caption{Schematic picture of the distribution functions of quarks and baryons in the quarkyonic saturation regime.  Saturated quarks in the blue shaded region in $n_{\mathrm{Q}}^{\text{bulk}}$ are dual to suppressed baryons in the red shaded region below $k_{\mathrm{F}}\simeq \Nc q_{\text{bulk}}$.  Quarks in $n_{\mathrm{Q}}^{\text{surf}}$ are pushed outward, which is contained in the corresponding baryon Fermi sea with the width $\Delta k$ that is decreasing as the density grows up.}
    \label{fig:schematic_dist}
\end{figure}

Once the critical density of quarkyonic saturation is reached, the quark Fermi sea begins to fill up, as illustrated in the left panel of Fig.~\ref{fig:schematic_dist}.  Above the saturation density, quarks fill the Fermi sea up to $q_{\mathrm{bulk}}$.  Since the density at which quarkyonic saturation occurs is parametrically at most of order $\Lambda^3$, the nucleons at this saturation are nonrelativistic with a Fermi momentum of at most order $\Lambda$, and a baryon chemical potential:
\begin{equation}
  \muB \;\sim\; m_N + \frac{\Lambda^2}{2m_N} \;\sim\; m_N(1 + \mathcal{O}(\Nc^{-2})) \,.
\end{equation}
The onset chemical potential for the transition to Quarkyonic Matter is very close to the baryon mass in the large-$\Nc$ limit.  The quarkyonic saturation of the quark distribution function drives baryons into the relativistic regime, since the induced change in the chemical potential is of $\mathcal{O}(\Nc)$ as explained below.

After the transition to Quarkyonic Matter, $q_{\mathrm{bulk}}$ in $f_{\mathrm{Q}}$ develops, and ordinary nuclear matter seen in $f_{\mathrm{B}}$ sits on the Fermi surface with the occupation number one, surrounding a Fermi sea up to $k_{\mathrm{F}}\simeq \Nc\,q_{\mathrm{bulk}}$ in $f_{\mathrm{B}}$, as summarized in Fig.~\ref{fig:schematic_dist}.
As the density increases, the Fermi sea of quarks becomes larger.  In terms of the baryon distribution, a fully occupied quark Fermi sea with occupation number one up to the momentum $q_{\mathrm{bulk}}$ (a shaded region in the left panel of Fig.~\ref{fig:schematic_dist}) corresponds to an under-occupied baryon Fermi sea with the occupation number suppressed by $\Nc^{-3}$ up to the momentum $k_{\mathrm{F}}$ (a shaded region in the right panel of Fig.~\ref{fig:schematic_dist})~\cite{Fujimoto:2023mzy}.  This can be understood from Eq.~\eqref{eq:idylliq}; if $q_{\mathrm{bulk}} \gg \Lambda$ and $f_{\mathrm{Q}}(\bq)=1$ in the saturated regime, we can make an approximation of $K_N(\bq-\bk/\Nc)\simeq (2\pi)^3\delta^{(3)}(\bq-\bk/\Nc)$, which results in $f_{\mathrm{Q}}(\bq) = 1 \simeq \Nc^3 f_{\mathrm{B}}(\Nc \bq)$ for $|\bq|\le q_{\mathrm{bulk}}$.

From this picture of Quarkyonic Matter, we understand that the EoS becomes stiffened once quarkyonic saturation occurs. After the phase-space integration of the distribution functions, the quark number density has two contributions:
\begin{equation}
    n_{\mathrm{Q}} = n_{\mathrm{Q}}^{\text{bulk}} + n_{\mathrm{Q}}^{\text{surf}}\,,
\end{equation}
where $n_{\mathrm{Q}}^{\text{bulk}}$ is the contribution of the degenerate quark Fermi sea up to $q_{\mathrm{bulk}}$.  At the density of quarkyonic saturation, there is only surface contribution that is $n_{\mathrm{Q}}^{\text{surf}} = \Nc n_{\mathrm{B}}^{\text{sat}}$ at that point, and this approximately holds as long as $n_{\mathrm{Q}}^{\text{surf}} \gg n_{\mathrm{Q}}^{\text{bulk}}$; see the left panel of Fig.~\ref{fig:schematic_dist}.  The total baryon number chemical potential is approximately
\begin{equation}
    \muB = \Nc\, \muQ + \delta\muB^\mathrm{sat}\,,
    \label{eq:chem_saturation}
\end{equation}
where $\muQ$ is the quark chemical potential corresponding to the Fermi momentum $q_{\mathrm{bulk}}$ and $\delta\muB^\mathrm{sat}$ is the incremental baryon chemical potential corresponding to the nucleons on the Fermi surface, as indicated in the right panel of Fig.~\ref{fig:schematic_dist}.  In this model, these nucleons form a shell of occupation number one in $f_{\mathrm{B}}$.  For the exponential kernel, the thickness of the shell in momentum, $\Delta k$ (denoted by $\Delta_{\mathrm{B}}$ in Ref.~\cite{Fujimoto:2023mzy}), decreases with density and approaches $\Lambda/\Nc^2$ at high density~\cite{Fujimoto:2023mzy}, and $\delta\muB^{\mathrm{sat}} \simeq (k_{\mathrm{F}}/E_{k_{\mathrm{F}}})\,\Delta k$ with $k_{\mathrm{F}} \simeq \Nc q_{\mathrm{bulk}}$ is its energy width.  In the quarkyonic saturation regime, the total baryon number changes by $\sim \mathcal{O}(1)$ with increasing $\Nc\,\muQ$.  Therefore, $\muB$ changes by $\sim \mathcal{O}(\Nc)$ in response to adding the baryon number, and the system has become relativistic.  The sound velocity becomes enhanced because in the following expression of the sound velocity at $T=0$,
\begin{equation}
  v_s^2 = \frac{dp/d\muB}{d\varepsilon/d\muB} = \frac{n_{\mathrm{B}}/ \muB}{ d n_{\mathrm{B}} / d\muB } \,,
  \label{eq:vs2muB}
\end{equation}
the denominator, $d n_{\mathrm{B}}/d\muB$, is suppressed by $\Nc^{-1}$~\cite{McLerran:2018hbz}.  This transition to Quarkyonic Matter is the opposite of a conventional picture of the first-order phase transition from nuclear to quark matter.

\begin{figure}
    \centering
    \includegraphics[width=0.85\textwidth]{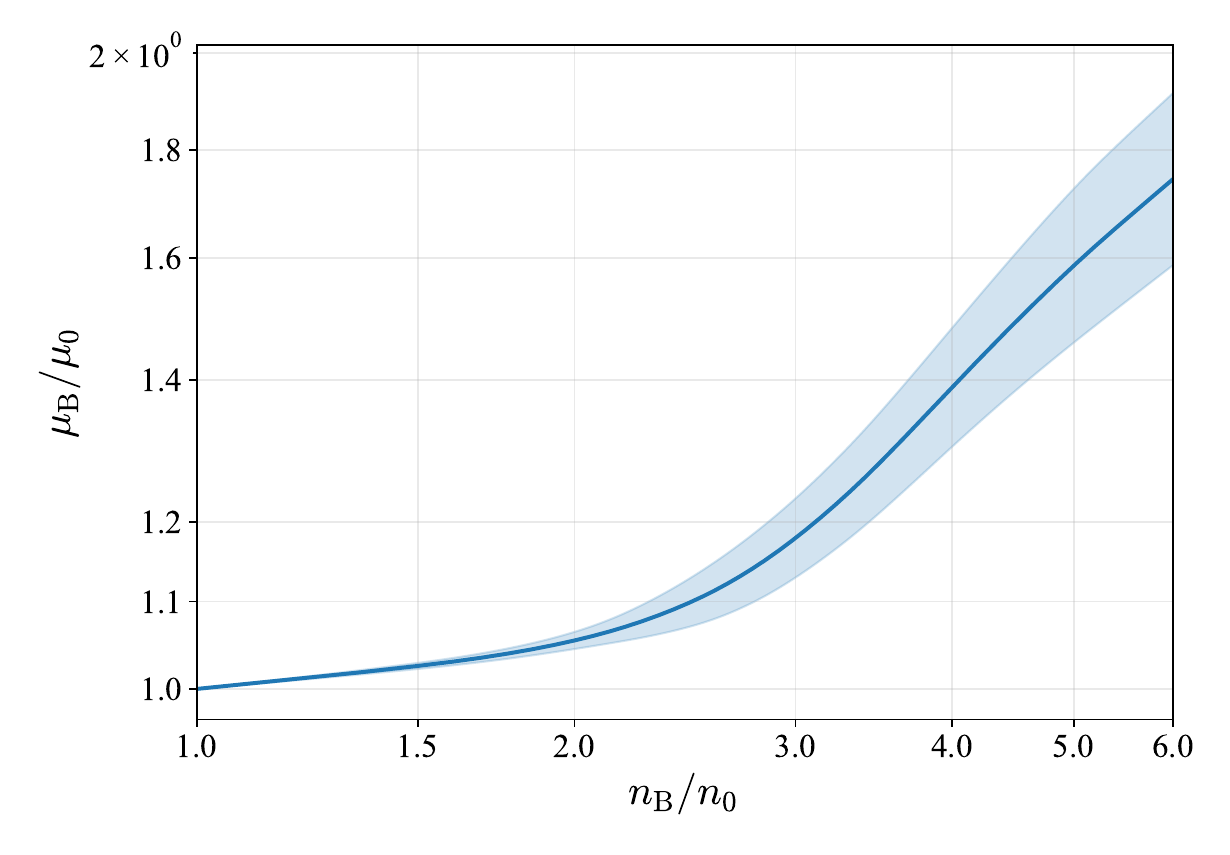}
    \caption{Baryon chemical potential vs.\ baryon number density, which we reconstruct from the observationally inferred EoS of neutron star matter; $\nB$ and $\muB$ from numerically integrated Eq.~\eqref{eq:muBsolved} using $N=8$ (density segment number) data of $v_s^2$ in Ref.~\cite{Fujimoto:2024cyv}.  The normalization, $n_0$, is the normal nuclear density and $\mu_0$ is the corresponding baryon chemical potential.  The quarkyonic saturation onset is seen in the density range $\nB/n_0= 2\text{-}3$.}
    \label{fig:muBnB}
\end{figure}

A more straightforward check of quarkyonic saturation is possible from the neutron star data of $\muB$ and $\nB$.  Here, we note that it is straightforward to solve $\muB$ from Eq.~\eqref{eq:vs2muB} as
\begin{equation}
    \frac{\muB}{\mu_0} = \exp\biggl[ \int_{n_0}^{\nB} \frac{v_s^2(n)}{n}\, dn \biggr]\,.
    \label{eq:muBsolved}
\end{equation}
Figure~\ref{fig:muBnB} shows the normalized $\muB$ as a function of the normalized $\nB$ based on the neutron-star inferred data~\cite{Fujimoto:2024cyv}.  We note at a glance that there is a transitional change around the density, $\nB/n_0 = 2\text{-}3$, above which $\muB$ grows rapidly in accordance with Eq.~\eqref{eq:chem_saturation}.  Thus, more clearly than the speed of sound peak, the transitional behavior in Fig.~\ref{fig:muBnB} indicates the onset of quarkyonic saturation.

\subsection{BEC-BCS Crossover}
\label{sec:bec-bcs}

The characterization of Quarkyonic Matter is associated with whether the bound states are good physical degrees of freedom.  A similar argument is found in the context of the crossover between the BCS (Bardeen--Cooper--Schrieffer) and BEC (Bose--Einstein condensate) regimes.
An attractively interacting Fermi gas forms pairs that condense into a superfluid, and within this superfluid phase there are two typical regimes.
The pair size is given by the coherence length, $\xi$, and the system has another characteristic scale, i.e., the interparticle distance $\sim q_{\mathrm{F}}^{-1}$.
When the interaction is weak, the gap is small, and the pair is formed by two particles only near the Fermi surface.  Since the momentum width cannot be large with a small gap, the corresponding pair size is large.  The weak-coupling regime with $\xi\gg q_{\mathrm{F}}^{-1}$ is called the BCS regime.

When the interaction is strong, in contrast, the pair becomes a molecular bound state smaller than the interparticle distance, and its Bose condensation causes superfluidity; this situation with $\xi \ll q_{\mathrm{F}}^{-1}$ is called the BEC regime. Since no symmetry distinguishes the BCS and BEC regimes, they can be connected with not a phase transition but a smooth crossover~\cite{Nozieres:1985zz, Randeria:2014}. In cold atomic systems, this crossover is experimentally realized by externally tuning the scattering length through a Feshbach resonance~\cite{Chin:2010crf}.

In gauge theories at finite density, the coupling strength is not a tunable parameter; instead, the density changes the effective coupling and the coherence length.  A theoretically reasonable scenario is the following:  The zero-density or empty vacuum in QCD and QCD-like theories is a strongly-coupled system, so when the superfluid transition occurs at some density, it is likely that the system is still in the nonperturbative regime.  Therefore, the pairs near the superfluid transition may well be tightly bound molecular states, that is, the pairs form a BEC-like condensate.  With increasing density, the effective coupling of QCD and QCD-like theories near the Fermi surface weakens due to asymptotic freedom.  Thus, BCS-like behavior of the pair condensate is expected to set in.  The sequential changes of ``the zero-density vacuum $\rightarrow$ BEC $\rightarrow$ BCS'' are thus driven by the density~\cite{Son:2000xc, Son:2000by, Hands:2006ve}.  In this picture, only two-body pairs (i.e., dimers) are involved.  For three-body bound states, such as baryons in three-color QCD, the corresponding description requires three-body fluctuations (trimers) to be treated explicitly~\cite{Tajima:2024qzj, Tajima:2026ukj, Tajima:2022iqw}.

In two-color QCD at finite baryon density and in three-color QCD at finite isospin density, the superfluid transition is driven by the diquark condensate and the pion condensate, respectively.  The transition occurs at the chemical potential set by the mass of the light pseudoscalar excitation (i.e., $\muQ \simeq m_{\mathrm{PS}}/2$ for dense two-color QCD~\cite{Kogut:2000ek} and $\muI = m_{\pi}$ at finite isospin density~\cite{Son:2000xc, Son:2000by}). In both cases, the superfluidity is associated with the breaking of global symmetry not coupled to the gauge fields.  This is therefore different from color superconductivity in QCD in which gauge fields are Meissner screened~\cite{Alford:2007xm}.

\begin{figure}
    \centering
    \includegraphics[width=0.9\textwidth]{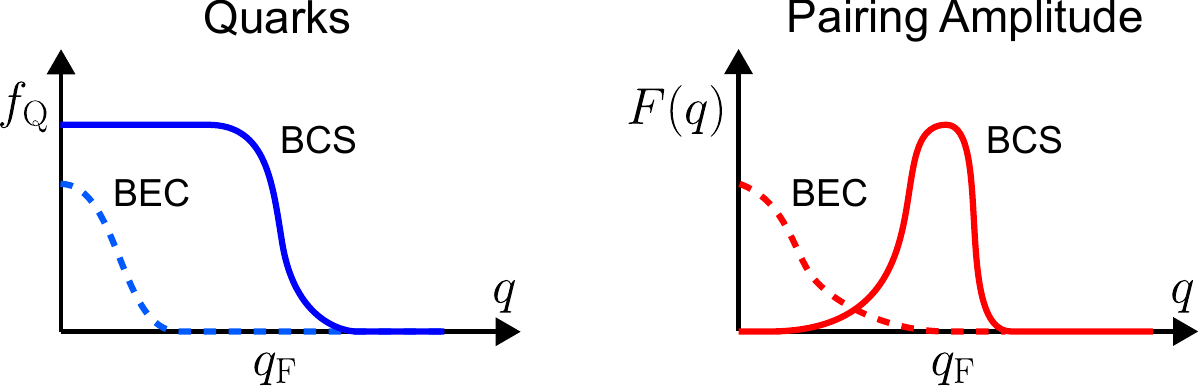}
    \caption{Schematic picture of the distribution functions of quarks and the pairing amplitudes corresponding to the BEC and BCS regimes.  In the figure, $q_{\mathrm{F}}$ stands for the quark Fermi momentum, while the distribution function in the BCS regime is smeared by the gap energy.  The vertical axis is in arbitrary units.}
    \label{fig:schematic_dist_BEC}
\end{figure}

Interestingly, using the quark distribution function, we can understand the BEC-BCS crossover in a way similar to Fig.~\ref{fig:schematic_dist}.  For this purpose of analogous discussion, let us consider the pairing amplitude in momentum space, $F(\bq) \sim \langle q(-\bq) q(\bq) \rangle$ (where the net momentum is taken to be vanishing for the BCS pairing), so that its momentum integral is the diquark condensate:
\begin{equation}
    \langle qq\rangle = \int\frac{d^3 q}{(2\pi)^3}\, F(\bq)\,.
\end{equation}
Then, in the BCS regime where the quasi-free quarks are relevant degrees of freedom, the following mean-field relation approximately holds (see the model calculation and mean-field thermodynamics in Ref.~\cite{Kojo:2021hqh}):
\begin{equation}
    |F(\bq)|^2 \simeq f_{\mathrm{Q}}(\bq) \bigl[ 1 - f_{\mathrm{Q}}(\bq) \bigr] \,,
\end{equation}
from which we conclude that nonzero $|F(\bq)|$ is localized around the quark Fermi momentum, $q_{\mathrm{F}}$, if $f_{\mathrm{Q}}(\bq)$ develops a Fermi sea; see Fig.~\ref{fig:schematic_dist_BEC}.  In the BEC side, in the case at low density, the quark distributions are localized in the momentum scale corresponding to the spatial extent of the molecular states, as indicated by the dashed curves in Fig.~\ref{fig:schematic_dist_BEC}.

Although the analogous behavior of the quark distribution is suggestive, the classification of the BCS regime in this scenario is based on the behavior of the quarks in the bulk, namely on the picture of free or weakly interacting quarks which dominate physical quantities such as the quark number density and the diquark condensate. 
The long-distance behavior of the gluonic degrees of freedom should be examined independently~\cite{Hands:2006ve, Braguta:2016cpw, Begun:2022bxj}.  If the gluonic sector retains a nonperturbative confining character, the BCS regime can be identified as a state in which the quarks behave as quasi-free degrees of freedom in bulk thermodynamics, while the gluons still impose confinement on excitations.  This is one possible manifestation of Quarkyonic Matter.

\subsection{Possible Realizations}

We will attempt to generalize these considerations about Quarkyonic Matter and related objects to theoretically idealized systems, as discussed below.
\vspace{1em}

\noindent
--- \textbf{Low-dimensional System} ---

We can immediately apply the above arguments to systems in $(d+1)$ spacetime dimensions, particularly a low-dimensional system with $(1+1)$ dimensions.  The essential modification is that the baryon occupation in the quarkyonic regime is suppressed not by $\Nc^{-3}$ but by $\Nc^{-1}$ (i.e., $\Nc^{-d}$ in general).  Also, the sound velocity at asymptotic densities, that is, the conformal limit of the sound velocity squared is $v_s^2 \rightarrow 1$ (i.e., $1/d$ in general) instead of $1/3$ known in $(3+1)$-dimensional systems.

QCD in $(1+1)$ dimensions provides a useful setting for Quarkyonic Matter, since excitations are confined in the vacuum and confinement persists at high density, while the bulk thermodynamics becomes quark-like~\cite{Kojo:2011en}.
Solvable models such as the massless two-flavor Schwinger model at finite density also allow explicit calculations of quark distributions, providing analytical insight into the convolution picture discussed above; see Appendix~\ref{app:schwinger2f}.

In the same way as two-color QCD in $(3+1)$ dimensions listed below, the BEC-BCS crossover discussed in Sec.~\ref{sec:bec-bcs} could change the structure of quark pairs within a single phase having a nontrivial diquark condensate.  In its $(1+1)$-dimensional counterpart, bound diquarks can still exist, but quantum fluctuations prevent their condensation in the thermodynamic limit, even at zero temperature.  The finite-density state of the single-flavor theory is instead consistent with a Tomonaga--Luttinger liquid, in which diquark correlations decay algebraically at large distances~\cite{Fujikura:2026tqk}.  Thus, the absence of a condensate would not rule out a BEC-BCS-like evolution of the pairs.  Establishing such an interpretation nevertheless requires information about the pair structure, beyond the quark-like thermodynamics as discussed below.

As mentioned in Sec.~\ref{sec:background}, gauge interactions gap the color degrees of freedom in $(1+1)$-dimensional $\mathrm{SU}(\Nc)$ QCD with one light quark flavor, and the low-energy sector at finite density is a Tomonaga--Luttinger liquid of color-singlet baryon-density fluctuations~\cite{Lajer:2021kcz}.  In the $(1+1)$-dimensional theory with $\Nf=1$ and $\Nc=2$, this bosonic density mode can coexist with a quark Fermi sea and quasi-free-quark bulk thermodynamics, exemplifying the quarkyonic picture~\cite{Fujikura:2026tqk}.
\vspace{1em}

\noindent
--- \textbf{Bose Condensation with Even $\Nc$} ---

Another straightforward generalization is to consider systems with even $\Nc$, for which the baryons are bosons.  The essential difference is that at low densities the baryon number is dominated by a Bose-condensed mode, i.e., Eq.~\eqref{eq:fB_Theta} is replaced with
\begin{equation}
  f_{\mathrm{B}}(\bk) = (2\pi)^3 \delta^{(3)}(\bk) \, n_\mathrm{cond}\,.
  \label{eq:fB_delta}
\end{equation}
Here, $n_\mathrm{cond}$ is the density of condensed baryons.  There is a critical density, $n_\mathrm{cond}^{\text{sat}}$, for which the KFC is satisfied, and the delta function solution in Eq.~\eqref{eq:fB_delta} is no longer valid for $n_\mathrm{cond} > n_\mathrm{cond}^\text{sat}$. 
The quark saturation mechanism in such a system has been examined, and the sound velocity is found to develop a peak~\cite{Kojo:2021hqh}.
Beyond this point, the momentum space distribution of bosonic baryons can extend to the Fermi surface energy.  Whether the condensate then carries a net momentum depends on how the constituent quarks are paired.  For the Cooper pair in the BCS state, the net momentum $\bk_{\mathrm{net}}$ of the condensate remains zero, while the FFLO (Fulde--Ferrell--Larkin--Ovchinnikov) state develops a condensate at $\bk_{\mathrm{net}}\neq \bm{0}$~\cite{Fulde:1964zz, Larkin:1964wok}.  In fact, it has been argued that high-density matter exhibits a pseudo-one-dimensional nature and the quarkyonic chiral spiral can be stabilized with $|\bk_{\mathrm{net}}|=2q_{\mathrm{F}}$~\cite{Kojo:2009ha}, where $q_{\mathrm{F}}$ represents the quark Fermi momentum.  This quarkyonic spiral state is similar to the FF state, as is the dual chiral density wave state~\cite{Nakano:2004cd}, while the genuine density modulation occurs in the solitonic state~\cite{Nickel:2009ke} as well as the LO state~\cite{Larkin:1964wok}.

Let us consider the quark sector.
Both the quarks contributing to the condensate and the surface part could sit on the Fermi surface, in analogy to the case with odd $\Nc$ with fermionic baryons.  Beyond the critical density, a filled Fermi sea of quarks is formed, corresponding to an under-occupied baryon number distribution with suppression by $\Nc^{-3}$ for three spatial dimensions and $\Nc^{-1}$ for one spatial dimension.  In the one-dimensional case, the bosonic baryons cannot form a condensate, strictly speaking, because infrared fluctuations destroy the order~\cite{Mermin:1966fe, Hohenberg:1967zz, Coleman:1973ci}.  Yet, the baryon number distribution may initially be concentrated at small momenta.  The essential difference between even and odd $\Nc$ is the degree of localization of the baryon number contribution at the Fermi surface.  Indeed, for odd $\Nc$, the surface contribution of nucleons is localized in a width of order $\Lambda/\Nc^2$~\cite{Fujimoto:2023mzy} at a Fermi energy $E \gtrsim \Nc\Lambda$ if the density is high enough.  For most practical purposes, at sufficiently high densities, the Fermi surface part may be regarded as a delta-function contribution.  This partially realizes qualitative continuity between even and odd $\Nc$.

Although, in principle, a Bose condensate of baryons may increase the sound velocity up to one, the transition to Quarkyonic Matter takes place when the bosons are still nonrelativistic.  There, the sound velocity is small, and the transition to Quarkyonic Matter can generate a large change upward for the sound velocity~\cite{Kojo:2021hqh}.
\vspace{1em}

\noindent
--- \textbf{Pion Condensation at High Isospin Density} ---

These considerations may also be generalized to systems at finite isospin chemical potential, $\muI$, for $\Nf=2$~\cite{McLerran:2026owy}.  The argument about screening is parallel to the previous case at high baryon density; thus we expect quarkyonic behavior in the window of $\LQCD \ll \muI \ll \sqrt{\Nc}\,\LQCD$.
In what follows, we consider finite isospin density and choose the sign convention so that positively charged pions form a condensate.  These pions are made up of $u$ and $\bar{d}$ quarks.  The counterpart of Eq.~\eqref{eq:idylliq} for mesons is
\begin{equation}
  f_{\mathrm{Q}}(\bq) = {1 \over {2\Nc}} \int {{d^3k} \over {(2\pi)^3}} K_{\mathrm{M}}(\bq - \bk/2) f_{\mathrm{M}}(\bk) \,,
  \label{eq:idylliq_meson}
\end{equation}
where $K_{\mathrm{M}}$ is the normalized probability of finding a quark inside a pion.
This expression needs an explanation.
First of all, $f_{\mathrm{M}}(\bk)$ is defined by the above expression;  see discussions below.
Next, regarding the convention, a baryon contains one quark of each color, whereas the single $u$ quark in the $\pi^+$ is in an equal superposition of all colors.  Since the $\pi^+$ is also a spin singlet, the same holds for the spin.  Then, the probability of finding a $u$ quark with a specific color and spin should be $1/(2\Nc)$, while $K_{\mathrm{M}}(\bq)$ is normalized to unity.  This is the reason why this overall factor is required in Eq.~\eqref{eq:idylliq_meson}.  Roughly speaking, as compared to the baryons, the mesons contain $\Nc^{-1}$ fewer quarks per color.  Indeed, if we perform the $\bq$ integration, we have
\begin{equation}
    \frac{n_{\mathrm{Q}}}{4\Nc} = \frac{n_{\mathrm{I}}}{2\Nc} \,,
    \qquad
    n_{\mathrm{I}} = \int \frac{d^3 k}{(2\pi)^3}\, f_{\mathrm{M}}(\bk)\,.
\end{equation}
Here, $n_{\mathrm{Q}}$ is the flavor-summed quark density defined in the same way as in Eq.~\eqref{eq:nQB}, with $u$ and $\bar{d}$ playing the role of the two flavors.  This gives $n_{\mathrm{Q}} = n_u + n_{\bar{d}} = 2\nI$, consistent with Eq.~\eqref{eq:nB_equiv} and with $\nI = \tfrac{1}{2}(n_u + n_{\bar{d}})$ in the standard convention.  

\begin{figure}
    \centering
    \includegraphics[width=0.9\textwidth]{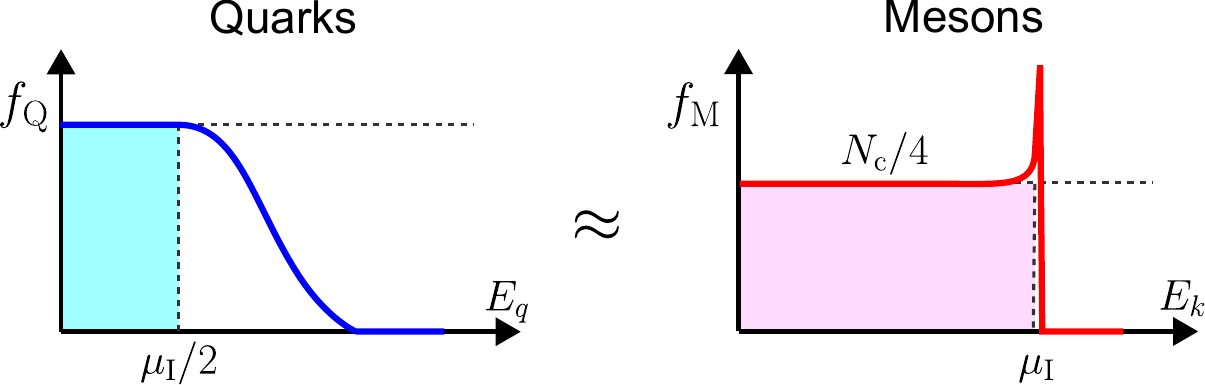}
    \caption{Schematic picture of the distribution functions of quarks and mesons in the quarkyonic saturation regime.}
    \label{fig:schematic_dist_meson}
\end{figure}

At low isospin density, the pion distribution is initially a delta function in momentum space, corresponding to the BEC formation.  Once quarkyonic saturation sets in above the KFC isospin density, pions begin to sit at a Fermi energy; see Fig.~\ref{fig:schematic_dist_meson}.  In the same way as $f_{\mathrm{B}}$, as mentioned above, $f_{\mathrm{M}}$ is defined through Eq.~\eqref{eq:idylliq_meson} with a fixed $K_{\mathrm{M}}$~\cite{Kojo:2021hqh}.  We note that the baryon model is based on the large-$\Nc$ picture in which the baryon momentum fraction is assumed to be shared by all valence quarks.  However, the meson is a system of a quark and an antiquark, and the net momentum of the meson and the quark--antiquark relative momenta can behave differently.  Nevertheless, the energy can still be shared by these constituents, which explains why we replace the horizontal axis in Fig.~\ref{fig:schematic_dist_meson} with the energy and the chemical potential, instead of the momentum and the Fermi momentum.
Indeed, the relative momenta could spread widely around zero or only locally near the Fermi momentum (see Fig.~\ref{fig:schematic_dist_BEC}), depending upon whether the condensate forms in the BEC/BCS regimes, and/or with the quarkyonic chiral spiral.

The critical isospin density for quarkyonic saturation is $n_{\mathrm{I}}^{\text{sat}}/\Lambda^3  \sim \mathcal{O}(\Nc)$ due to the factor, $1/(2\Nc)$, in Eq.~\eqref{eq:idylliq_meson}.
Therefore, since $\nI \propto \Nc \Lambda^2 \muI$ from chiral perturbation theory, where the squared pion decay constant is of order $\Nc\Lambda^2$ (the same estimate follows from the quark counting, $\nI \propto \Nc\muI^3$, at $\muI \sim \Lambda$), the critical value of the isospin chemical potential is $\muI^{\text{sat}} \sim \Lambda$.  This means that quarkyonic saturation needs the pion density to be parametrically large compared to $\Lambda^3$, and that the pions are relativistic at the quarkyonic transition density.

We shall further quantify the KFC in this case.
For $f_{\mathrm{M}}(\bk)=(2\pi)^3 \delta^{(3)}(\bk)  \nI$, it is easy to find that the KFC at $\bq=\bm{0}$ leads to
\begin{equation}
    1 = K_{\mathrm{M}}(0) \, \frac{\nI^{\text{sat}}}{2\Nc}\,,
    \label{eq:KFC_isospin}
\end{equation}
which makes a contrast to Eq.~\eqref{eq:KFC_QCD} for QCD at finite baryon density.
For $\nI > \nI^{\text{sat}}$, the pion distribution behaves similarly to the baryon number case in this model.  However, there is a significant difference.  Assuming uniform distributions for saturated quarks and mesons, for $q_{\mathrm{bulk}} \gg \Lambda$, we can approximately solve Eq.~\eqref{eq:idylliq_meson} as $f_{\mathrm{Q}}(\bq) = 1 \simeq (4/\Nc) f_{\mathrm{M}}(2\bq)$ for $|\bq| \le q_{\mathrm{bulk}}$.  The pion occupation number in this model, $\Nc/4$, reflects the $2\Nc$ color-spin states of the quarks to be filled.  It is large at large $\Nc$, in contrast to $\Nc^{-3}$ for baryons.  We could plot the momentum distribution functions as we did for the baryons in Fig.~\ref{fig:schematic_dist}, but the fate of the condensate originally sitting at $\bk=\bm{0}$ depends on the scenario.  Therefore, it would be more convenient to draw the distribution functions in terms of the energies, $E_q=\sqrt{\bq^2+m_{\mathrm{Q}}^2}$ and $E_k=\sqrt{\bk^2+m_\pi^2}$, as we mentioned above.  By introducing $\mu_{\mathrm{bulk}}=\sqrt{q_{\mathrm{bulk}}^2 + m_{\mathrm{Q}}^2}$, the corresponding pion or isospin chemical potential is $\mu_{\mathrm{I}} \approx 2\mu_{\mathrm{bulk}}$ for sufficiently light pions.  It would be a physically reasonable scenario that the condensate is pushed up to $\muI$ as a result of the quarkyonic saturation at low energies.  This situation is schematically illustrated in Fig.~\ref{fig:schematic_dist_meson}.

The sound velocity when the quarkyonic transition begins should be $\sim \mathcal{O}(1)$.  The quarkyonic saturated regime remains until $\muI \sim \sqrt{\Nc}\,\LQCD$, that is, the upper quarkyonic density is $\nI^{\text{cri}}/\Lambda^3 \sim \Nc\muI^3/\Lambda^3 \sim \mathcal{O}(\Nc^{5/2})$, so that $\nI^{\text{cri}} / \nI^{\text{sat}} \sim \mathcal{O}(\Nc^{3/2})$, corresponding to a range of isospin chemical potential, $\muI^{\text{cri}} / \muI^{\text{sat}} \sim \mathcal{O}(\Nc^{1/2})$.  Near the threshold of quarkyonic saturation, the isospin density is
\begin{equation}
    \nI = n_{\mathrm{I,Q}} + \nI^{\text{sat}}\,,
\end{equation}
and $n_{\mathrm{I,Q}}$ reaches a value of order $\nI^{\text{sat}}$ when the quark (isospin) chemical potential is $\muQ \sim \Lambda$.  For massless pions, the sound velocity in the state with nonzero condensate is already $v_s^2 \sim  1$.  The sound velocity at the transition to Quarkyonic Matter should be understood by the mechanism of filling the Fermi sea with quarks up to a density of order $\nI^{\text{sat}}$.  This means that the sound velocity change is of order one, while this is probably concealed within the dynamics of evolving meson density at the Fermi surface.  This is plausible because the pion condensation already arises without Quarkyonic Matter, and the sound velocity can become close to one.  Nevertheless, as the density gets larger, the system will rapidly approach that of an ideal gas of quarks, in spite of the fact that the confined phase still persists.  There, the sound velocity squared will drop down towards the conformal value of $1/3$ and would not stay near one.
In fact, the behavior of matter at finite isospin density has been
studied in Ref.~\cite{Chiba:2023ftg}, where a peak of the sound velocity
is obtained.

\section{Quark Degrees of Freedom in Lattice Simulation Data}
\label{sec:lattice}

We give an overview of lattice simulation results in various physical systems and interpret high-density behavior as an onset of quark degrees of freedom.  We specifically consider the following sign-problem-free systems in order.

\subsection{\texorpdfstring{QC$_2$D$_2$ at $\bm{\muB\neq0}$}{QC2D2 at muB not eq 0}}

This part addresses $(1+1)$-dimensional two-color QCD at finite baryon density.  The low dimensionality allows for high-precision numerical simulations as well as analytical insights.  For $(1+1)$-dimensional $\mathrm{SU}(\Nc)$ QCD, analyses using bosonization and the Bethe ansatz have identified the low-energy sector at finite density as a Tomonaga--Luttinger liquid~\cite{Lajer:2021kcz}.  High-precision computations of $(1+1)$-dimensional QCD, denoted by QCD$_2$ here (or QC$_2$D$_2$ when the choice of $\Nc=2$ matters), have also been performed from first principles using matrix product states in the Hamiltonian lattice formulation~\cite{Hayata:2023pkw,Fujikura:2026tqk}.  In the single-flavor two-color theory, uniform matrix product states exploit translational invariance to extract bulk properties directly in the thermodynamic limit, without boundary effects~\cite{Fujikura:2026tqk}.  The lattice data analysed in this subsection were obtained with this method.

\subsubsection{Thermodynamic properties}

Two features distinguish QCD$_2$ from QCD$_4$ of our ultimate interest.  First, a relativistic gas in $(1+1)$ dimensions has sound speed $v_s^2 = 1$ rather than $1/3$, so this is the value that the sound speed should approach in the asymptotically high-density limit.  Second, the Debye screening is less sensitive to the chemical potential.  This is because the coupling constant $g$ in QCD$_2$ is dimensionful, i.e., $g \sim \mathrm{(mass)}$, which replaces the typical mass scale.  In fact, the one-loop contribution to the Debye mass is of order
\begin{equation}
  M_\mathrm{Debye}^2 \;\sim\; {\lambda \over \Nc} \,.
\end{equation}
Here, in the same way as QCD$_4$, the 't~Hooft coupling, $\lambda=g^2\Nc$, is introduced.
The Debye mass thus does not grow with $\muQ$.  At large $\Nc$ with $\lambda$ fixed, we see that $M_\mathrm{Debye}^2 \sim \mathcal{O}(\Nc^{-1}) \to 0$ for any $\muQ$, so the theory remains confined at arbitrarily high density.  Even for $\Nc = 2$ as we discuss below, the absence of the $\muQ$ dependence implies that finite density has little effect on screening.

A further feature of QCD$_2$ is that strong infrared fluctuations prohibit spontaneous breaking of continuous symmetry~\cite{Mermin:1966fe,Coleman:1973ci} and no Bose condensate forms.  Below, we focus on the special case of $\Nc=2$ and $\Nf=1$, i.e., QC$_2$D$_2$, and discuss the numerical results shown in Ref.~\cite{Fujikura:2026tqk}.  Therefore, although baryons are bosonic diquark states for $\Nc=2$, diquark condensation does not occur in QC$_2$D$_2$ in the thermodynamic limit.

\begin{figure}
  \centering
  \includegraphics[width=0.49\columnwidth]{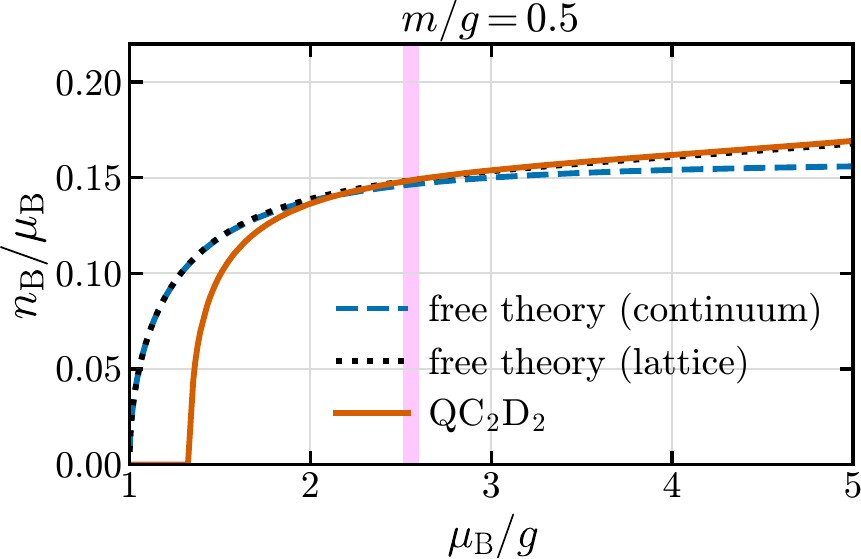}
  \hspace{0.2em}
  \includegraphics[width=0.49\columnwidth]{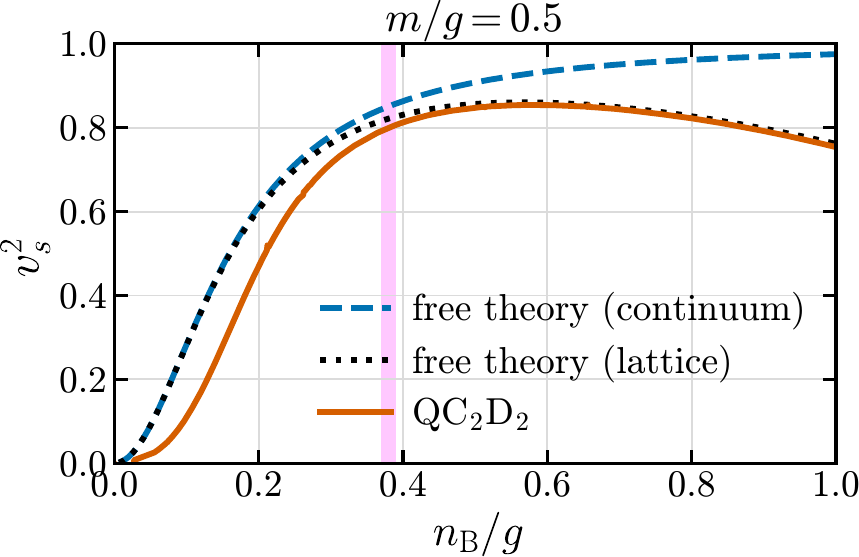}
  \caption{(Left) Ratio of the baryon number density to the chemical potential, $n_{\mathrm{B}}/\muB$, as a function of $\muB/g$ for quark mass $m/g=0.5$.  The crossing point between the QC$_2$D$_2$ data and the lattice free theory is indicated by a thick vertical line.  (Right)  Squared sound speed as a function of $n_{\mathrm{B}}/g$ for $m/g=0.5$~\cite{Fujikura:2026tqk}.}
  \label{fig:NBoverMuB_mg0p5}
\end{figure}

The left panel in Fig.~\ref{fig:NBoverMuB_mg0p5} shows the baryon number density as a function of dimensionless chemical potential in QC$_2$D$_2$.  Above threshold, the baryon density rises rapidly to the limit of a massive free-quark gas.  At large chemical potentials, $\muB/g\sim 5$, lattice artifacts become visible in the left panel in Fig.~\ref{fig:NBoverMuB_mg0p5}.  This is because the corresponding Fermi momentum in units of the unit cell $2a$ is $2\pi n_{\mathrm{B}}a \approx 1.3$, which is not small compared with unity, as required for the continuum description.  There, around $\muB/g \gtrsim 3$, the data (orange solid curve) track the free lattice theory (dotted curve) rather than the free continuum theory (blue dashed curve).  The QC$_2$D$_2$ data cross the lattice free theory at $\muB/g\approx 2.56$, which corresponds to $n_{\mathrm{B}}/\muB\approx 0.149$ and thus $n_{\mathrm{B}}/g\approx 0.380$, as indicated by vertical lines in Fig.~\ref{fig:NBoverMuB_mg0p5}.

In the right panel in Fig.~\ref{fig:NBoverMuB_mg0p5}, the squared sound speed, $v_s^2$, is plotted against the dimensionless baryon number density.  Again, it rises rapidly till $n_{\mathrm{B}}/g \lesssim 0.3$.  The sound speed exhibits a maximum, which is an artifact of the finite lattice spacing, as shown by the dotted curve representing the lattice free theory.

\subsubsection{Distribution functions}

We shall discuss the quark distribution function in momentum space.  At zero temperature, the fermionic number density is generally written as
\begin{equation}
    f_{\mathrm{Q}}(q) = \frac{1}{\Nc} \int_{-\infty}^{\infty} d\omega\, \bigl[\Theta(\muQ - \omega) - \Theta(-\omega)\bigr] \, \mathrm{tr}[\gamma^0 \rho_{\mathrm{Q}}(\omega,q)] \,.
    \label{eq:fQ_spect}
\end{equation}
Here, the trace runs over the Dirac and color indices, and the second term subtracts the vacuum contribution.  The spectral function is the imaginary part of the retarded Green's function: $\rho_{\mathrm{Q}}(q)=-\frac{1}{\pi}\mathrm{Im} D_{\mathrm{Q}}(q)$, where $D_{\mathrm{Q}}(q)$ is a gauge-invariant propagator with the Wilson line.  In the quasi-particle approximation, $D_{\mathrm{Q}}(q) \sim Z_{\mathrm{Q}}(q) q_\mu\gamma^\mu/[(\omega+i0^+)^2 - \varepsilon(q)^2]$, the spectral function is $\rho_{\mathrm{Q}}(q) \sim Z_{\mathrm{Q}}(\omega,q) \delta(\omega-\varepsilon(q))$, leading to $f_{\mathrm{Q}}(q) \sim Z_{\mathrm{Q}}(\varepsilon(q),q) \Theta(\muQ-\varepsilon(q))$.  Even in this simple approximation, thus, the distribution function has nontrivial momentum dependence due to the dressing function, $Z_{\mathrm{Q}}(\varepsilon(q),q)$.  In the strongly correlated system, the spectral function may not be dominated by a single pole, and it could have a continuous distribution.

The spectral discussion above gives a qualitative picture of how interactions modify the free Fermi distribution.  To obtain an explicit relation, we use the definition of $f_{\mathrm{Q}}$ as the Fourier transform of the vacuum-subtracted equal-time correlator.  For one massless quark flavor in the continuum $\mathrm{SU}(\Nc)$ gauge theory at zero temperature, Appendix~\ref{app:factorization} uses a spatial chiral rotation to factorize this correlator into its free counterpart and a factor determined by the zero-density theory.  Fourier transformation of this product gives
\begin{equation}
  f_{\mathrm{Q}}(q) = \int \frac{dq'}{2\pi}\, n_{\mathrm{F}}(q')\,\Kone(q-q') \,,
  \qquad
  n_{\mathrm{F}}(q') = \Theta(\muQ - |q'|) \,,
  \label{eq:main:conv}
\end{equation}
where $\Kone(q)$ is the Fourier transform of the zero-density factor and satisfies $\int\frac{dq}{2\pi} \Kone(q) = 1$.  The kernel's dependence on the momentum difference $q-q'$ and its independence of $\muB$ follow from this factorization, without requiring the explicit form of the zero-density correlator.

The kernel $\Kone(q)$ represents the quark distribution inside a single baryon.  To see this, let us fix the total baryon number to unity, i.e., $N_{\mathrm{B}}=L \nB=1$, where $L$ is the volume (the spatial length) of the system.  We shall then consider the limit of $L\to\infty$, so that $\nB$ goes to zero.  Since $f_{\mathrm{Q}}(q)$ is a phase-space distribution, the corresponding momentum distribution integrated over space is obtained by $f_{\mathrm{Q}}(q)L$.  For sufficiently small $\muQ \ll \Lambda$, where $\Lambda$ is the damping scale of $\Kone(q)$, we can approximately factorize the $q'$ integration as $\int\frac{dq'}{2\pi} n_{\mathrm{F}}(q')\Kone(q) = (\muQ/\pi) \Kone(q)$.  Noting $\nB = \muQ/\pi$, we find $f_{\mathrm{Q}}(q) \simeq \nB \Kone(q)$ from Eq.~\eqref{eq:main:conv}.  Therefore,
\begin{equation}
  \lim_{L\to\infty} f_{\mathrm{Q}}(q)L = \lim_{L\to \infty} L \nB \Kone(q) = \Kone(q) \,,
\end{equation}
where $N_{\mathrm{B}}=L \nB=1$ is used.

Equation~\eqref{eq:main:conv} has a structure similar to Eq.~\eqref{eq:idylliq} in the quarkyonic model.  However, since $n_{\mathrm{F}}(q')$ is the distribution of not baryons but quarks, there is no $1/\Nc$ factor as seen in the quarkyonic model.  Nevertheless, we can translate Eq.~\eqref{eq:main:conv} into a form that is manifestly the same as the quarkyonic model.  Introducing the baryon momentum $k=\Nc q'$ in Eq.~\eqref{eq:main:conv} and using $\Nc\muQ = \muB$, we find:
\begin{equation}
  f_{\mathrm{Q}}(q) = \int\frac{dk}{2\pi} \, \Kone(q - k/\Nc) \, f_{\mathrm{B}}(k)\,,\qquad
  f_{\mathrm{B}}(k) = \frac{1}{\Nc}\,\Theta(\muB-|k|)\,,
  \label{eq:main:convB}
\end{equation}
which is the one-dimensional analog of Eq.~\eqref{eq:idylliq}.  Interestingly, the suppression factor by $1/\Nc$ in $f_{\mathrm{B}}(k)$ is consistent with the prediction from the quarkyonic model once the quarkyonic saturation occurs.  However, it should be noted that Eq.~\eqref{eq:main:convB} always holds for any density;  $1/\Nc$ suppression is there even before the quarkyonic saturation as we see below, and the baryon Fermi surface as sketched in the right panel of Fig.~\ref{fig:schematic_dist} does not grow.

\subsubsection{Quarkyonic saturation}

\begin{figure*}
  \centering
  \begin{minipage}{0.48\linewidth}
    \centering
    \includegraphics[width=\linewidth]{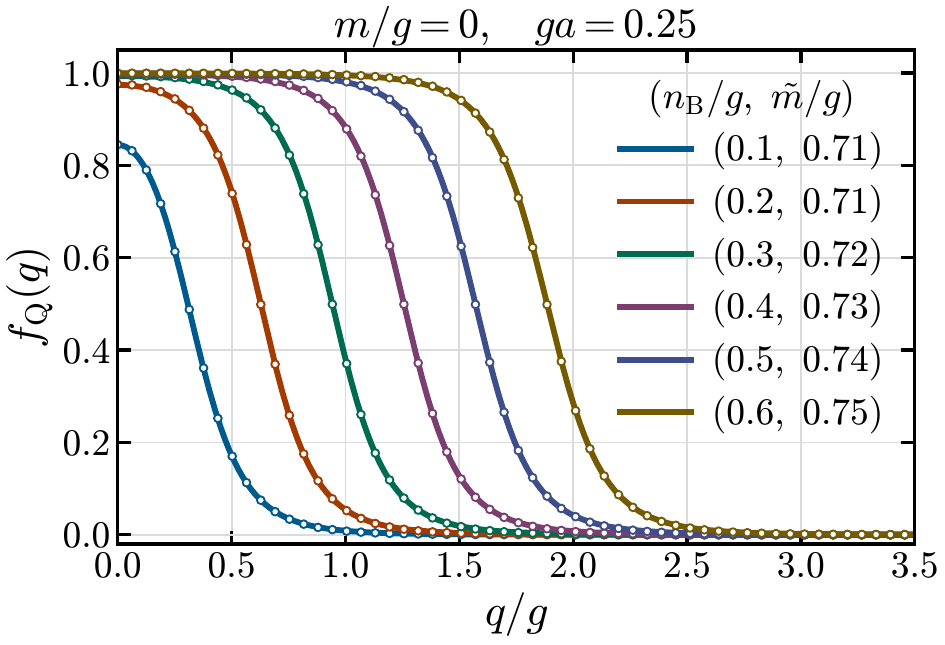}\\[-0.5ex]
    (a)
  \end{minipage}\hfill
  \begin{minipage}{0.48\linewidth}
    \centering
    \includegraphics[width=\linewidth]{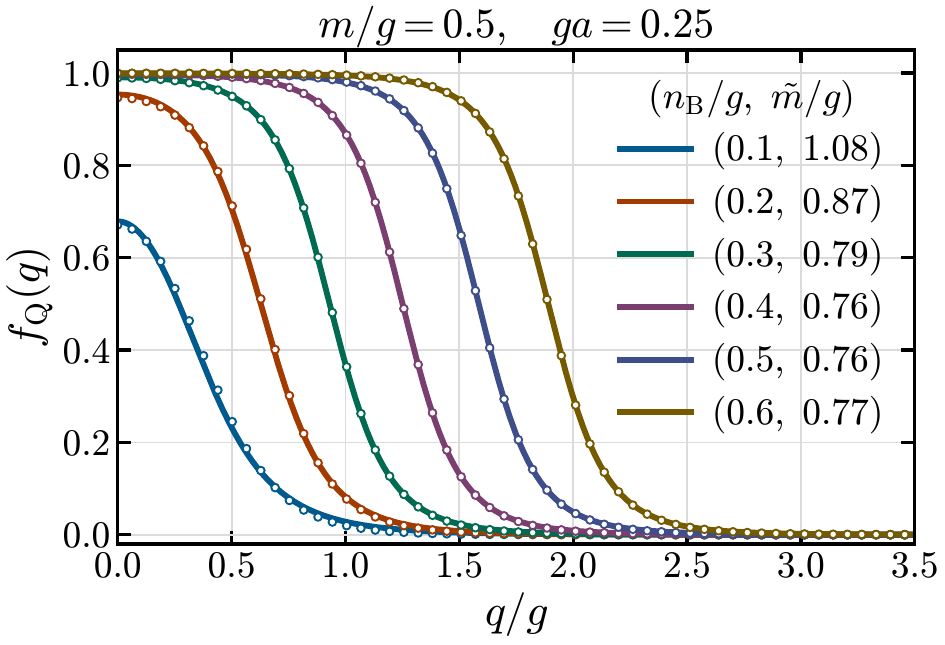}\\[-0.5ex]
    (b)
  \end{minipage}
  \caption{Quark distribution $f_{\mathrm{Q}}(q)$ at $ga=0.25$ for various baryon densities, $\nB/g=0.1,\dots, 0.6$.  The open circles show the lattice data and the solid curves represent the convolution-form fits with $\tilde{m}$ adjusted.  (a) Results in the massless case, $m/g=0$, to verify that $\tilde{m}$ effectively absorbed the difference between U(1) and SU(2). (b) Results in the massive case, $m/g=0.5$, to verify that the massless form is well applicable to the massive case with $\tilde{m}$ adjusted.}
  \label{fig:qdist_fitform}
\end{figure*}

Figure~\ref{fig:qdist_fitform} shows the quark distribution $f_{\mathrm{Q}}(q)$ in QC$_2$D$_2$ at six baryon densities, $n_{\mathrm{B}}/g\approx0.1,\dots,0.6$.  Panels (a) and (b) of Fig.~\ref{fig:qdist_fitform} present the results for $m/g=0$ and $0.5$, respectively.  The open circles are the lattice data at $ga=0.25$, and the curves are the one-parameter fits described below.

At low densities, the occupation at $q=0$ remains below unity, so a Fermi sea has not yet formed.  Only around $\nB^{\text{sat}}/g\approx 0.2$ does the occupation become close to unity, and a Fermi sea begins to develop, which signifies the quarkyonic saturation.  Even at higher densities, the distribution never acquires a sharp Fermi surface but remains smeared, which is naturally interpreted as confinement that persists for any density in QC$_2$D$_2$.

This behavior of $f_{\mathrm{Q}}(q)$ with increasing baryon density is completely consistent with that of the quarkyonic model~\cite{Fujimoto:2023mzy}; see the illustration in Fig.~\ref{fig:schematic_dist}.  It is a natural order that $n_{\mathrm{B}}^{\text{sat}}/g \approx 0.2 < 0.380$ where the thermodynamic properties are well described by a quasi-free gas of quarks for $n_{\mathrm{B}}/g \gtrsim 0.380$.

Let us explain the one-parameter fits shown by the curves in Fig.~\ref{fig:qdist_fitform}.  From the exact convolution form in Eq.~\eqref{eq:main:conv}, we can analytically compute $f_{\mathrm{Q}}(q)$ if the kernel, $\Kone(q)$, is obtained.  The shape of $\Kone(q)$ is not known in the closed form for SU($\Nc$), but it should be a reasonable approximation to employ its form, $K_{\text{U(1)}}$, from the exactly solvable two-flavor Schwinger model; see Appendix~\ref{app:schwinger2f} and Eqs.~\eqref{eq:app:D}--\eqref{eq:app:np}.  The idea is that the difference between U(1) and SU(2) can reasonably be adjusted by a single scaling parameter, $\tilde m$.

More specifically, we consider the massless two-flavor Schwinger model, in which the fermion flavors $u$ and $d$ have the same unit electric charge and chemical potentials $\mu_u=+\muI/2$ and $\mu_d=-\muI/2$, where $\muI$ is the isospin chemical potential.  At zero temperature, the resulting quark distribution integrates to the isospin density $n_I=\muI/(2\pi)$.  Appendix~\ref{app:schwinger2f} derives this distribution as the convolution of the free Fermi factor $\Theta(\muI/2-|q'|)$ with a $\muI$-independent kernel. This kernel has the exact integral representation
\begin{equation}
\begin{aligned}
  K_{\text{U(1)}}(q)
  &= \frac{2}{m_\gamma}\int_0^\infty du\,
     \cos\!\left(\frac{q}{m_\gamma}u\right)\mathcal D(u) , \\
  \mathcal D(u)
  &:= \exp\!\left\{\frac{1}{4}\left[
     2K_0(u)+2\ln\frac{u}{2}+2\gamma_E
     -\int_0^u dv\,(u-v)K_0(v)
     \right]\right\} .
\end{aligned}
  \label{eq:schwinger-kernel}
\end{equation}
Here, $u$ is dimensionless, and $m_\gamma=g_{\mathrm{U}(1)}\sqrt{2/\pi}$ is the gauge boson mass.  The function $K_0$ is the modified Bessel function of the second kind of order zero, and $\gamma_E$ is Euler's constant.  The value $\mathcal D(0)=1$ is defined by continuity.

The Schwinger-model kernel depends on $q/m_\gamma$ and has dimension inverse mass.  For the $\mathrm{SU}(\Nc)$ distributions at finite baryon density, we adopt this density-independent kernel shape as a modeling assumption and replace its mass scale by $\tilde m$.  The resulting one-parameter ansatz is
\begin{equation}
  \Kone(q;\tilde{m}) = \frac{m_\gamma}{\tilde m}\,
  K_{\text{U(1)}}\!\left(\frac{m_\gamma}{\tilde m}q\right) \,.
  \label{eq:kernel-rescaling}
\end{equation}
This rescaling preserves the normalization, $\int dq\,\Kone(q)/(2\pi)=1$.

\begin{figure}
  \centering
  \includegraphics[width=0.80\columnwidth]{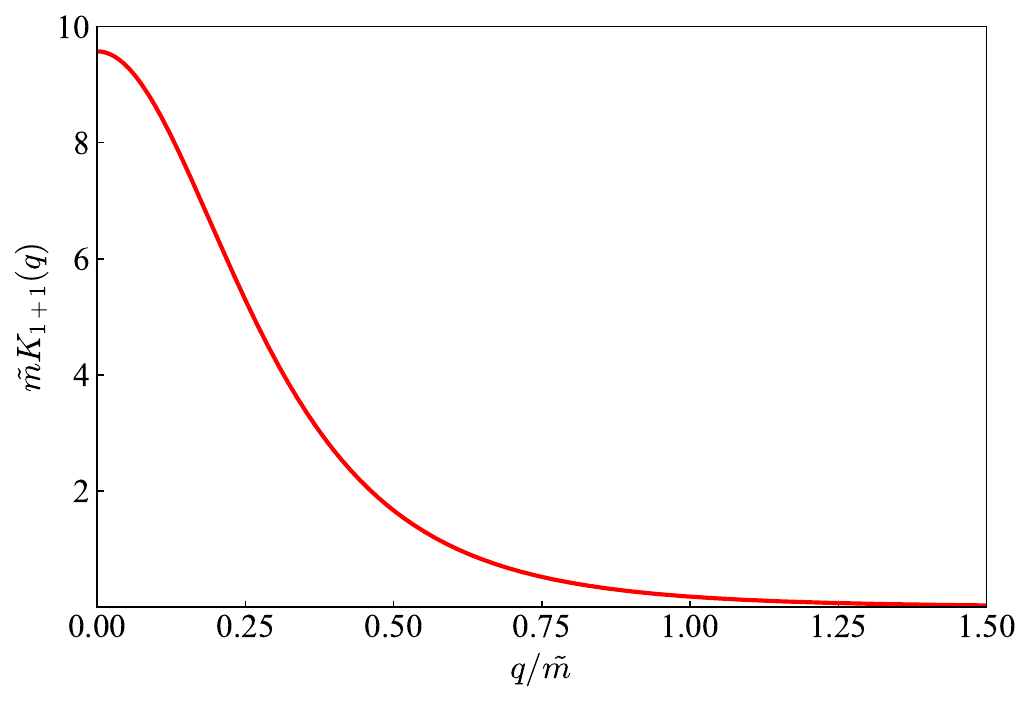}
  \caption{Density-independent smearing kernel $\tilde m \Kone(q)$ defined in Eq.~\eqref{eq:kernel-rescaling}, plotted against $q/\tilde m$.}
  \label{fig:kernel_convolution}
\end{figure}

The kernel $\Kone(q)$ is shown in Fig.~\ref{fig:kernel_convolution}.  The kernel is a smooth function of momentum, with a width of order the gauge-sector mass scale, $\tilde m/g\sim \mathcal{O}(1)$.  We use the convolution form of Eq.~\eqref{eq:main:conv} to fit the measured $m=0$ distributions.  For these fits, we take $n_{\mathrm{F}}(q')=\Theta(q_{\mathrm{F}}-|q'|)$ and use the kernel ansatz in Eq.~\eqref{eq:kernel-rescaling}.  Thus, the explicit form of the fitting function is
\begin{equation}
  f_{\mathrm{Q}}^{\mathrm{fit}}(q;n_{\mathrm{B}},\tilde m)
  = \int_{-q_{\mathrm{F}}}^{q_{\mathrm{F}}}\frac{dq'}{2\pi}\,\Kone(q-q';\tilde{m}) \,.
  \label{eq:quark-fit}
\end{equation}
We fix the overall coefficient to unity, set $q_{\mathrm{F}}=\pi n_{\mathrm{B}}$ from the measured density, and treat $\tilde m/g$ as a fitting parameter to minimize the squared deviations from the data.  For the SU(2) lattice data at $m/g=0$ in Fig.~\ref{fig:qdist_fitform}(a), the fitted scale increases slightly from $\tilde m/g=0.711$ at $n_{\mathrm{B}}/g\approx0.1$ to $0.753$ at $n_{\mathrm{B}}/g\approx0.6$.  The same fitting form works equally well for other lattice spacings and bond dimensions.

As a further empirical check, we apply the same functional form derived from the massless Schwinger model to the massive SU(2) lattice data at $m/g=0.5$ as shown in Fig.~\ref{fig:qdist_fitform}(b), again adjusting only the effective scale $\tilde m/g$ at each density and changing nothing else.  For the massive data, the fitted scale falls from $\tilde m/g=1.08$ at $n_{\mathrm{B}}/g\approx0.1$ to $0.769$ at $n_{\mathrm{B}}/g\approx0.6$.  Over the density range studied, the fitted scale for the massive SU(2) data approaches that for the massless ($m/g=0$) SU(2) data at the corresponding density.  At this latter density, the fitted scale agrees closely with the massless SU(2) value, $\tilde m/g=0.753$.  This density dependence is naturally understood.  The convolution form of Eq.~\eqref{eq:main:conv}, with a density-independent kernel, is derived only at $m=0$; see Appendix~\ref{app:factorization}.  At nonzero quark mass this factorization is no longer guaranteed, so the effective scale extracted from the fit can in general depend on the density.

The massless $(1+1)$-dimensional theory thus provides an example in which $f_{\mathrm{Q}}$ approaches saturation without a baryon momentum shell.
This absence is expected rather than paradoxical.
In the quarkyonic model, the shell evolves continuously from the baryon Fermi sea with $f_{\mathrm{B}}=1$ that exists before the saturation~\cite{Fujimoto:2023mzy}.
In the massless theory, the baryon occupation is $1/\Nc$ at all densities [see Eq.~\eqref{eq:main:convB}] because the baryons are massless and the density rises from $\muB=0$ without a threshold.
The growth of $f_{\mathrm{Q}}(0)$ at low density in Fig.~\ref{fig:qdist_fitform}(a) follows from $f_{\mathrm{Q}}\simeq n_{\mathrm{B}}\Kone$ and does not indicate a baryon Fermi sea with $f_{\mathrm{B}}=1$.
Hence, this example does not test whether a shell forms in $(3+1)$-dimensional matter of heavy baryons.
With a nonzero quark mass, a dilute baryonic regime exists, but $f_{\mathrm{Q}}$ alone does not determine $f_{\mathrm{B}}$; see the discussion of baryon correlations below.

\subsubsection{Baryon correlations}

The preceding analysis concerns quark observables.  The same lattice ground states also allow us to calculate baryonic correlations.  A gauge-invariant, parity-even continuum operator for the baryonic channel is
\begin{equation}
  \Delta_B(t,x) \coloneqq \ri\,\epsilon^{c_1c_2}
  \psi_{c_1}^{T}(t,x)C\gamma^5\psi_{c_2}(t,x) .
  \label{eq:diquark-operator}
\end{equation}
Here, $\psi_c$ is the continuum quark field, $c=1,2$, $\epsilon^{c_1c_2}$ is antisymmetric with $\epsilon^{12}=1$, $C$ is the charge-conjugation matrix, and $\gamma^5=\gamma^0\gamma^1$ in $(1+1)$ dimensions.  This operator has unit baryon number.

For the numerical results, we use the staggered fermion operator $\Delta_B(n)$ defined in Ref.~\cite{Fujikura:2026tqk}, where $n$ is the unit-cell label and $a$ is the lattice spacing.  The brackets below denote connected ground state expectation values at equal time, either at density $n_{\mathrm{B}}$ or in the vacuum.  We define $C_\Delta$ and its dimensionless structure factor $F_\Delta$ at diquark momentum $k$ by
\begin{equation}
\begin{aligned}
  C_\Delta(n)
  &\coloneqq \langle\Delta_B^\dagger(0)\Delta_B(n)\rangle_{n_{\mathrm{B}},\mathrm{conn}}
  -\langle\Delta_B^\dagger(0)\Delta_B(n)\rangle_{0,\mathrm{conn}} \,, \\
  F_\Delta(k/g)
  &\coloneqq \frac{1}{g^2}\sum_{n=-\infty}^{\infty}
  e^{\ri k(2an)}C_\Delta(n) \,,
  \qquad -\frac{\pi}{2ga}\leq \frac{k}{g}\leq\frac{\pi}{2ga} \,.
\end{aligned}
  \label{eq:diquark_two-point_function}
\end{equation}

\begin{figure*}
  \centering
  \includegraphics[width=0.85\textwidth]{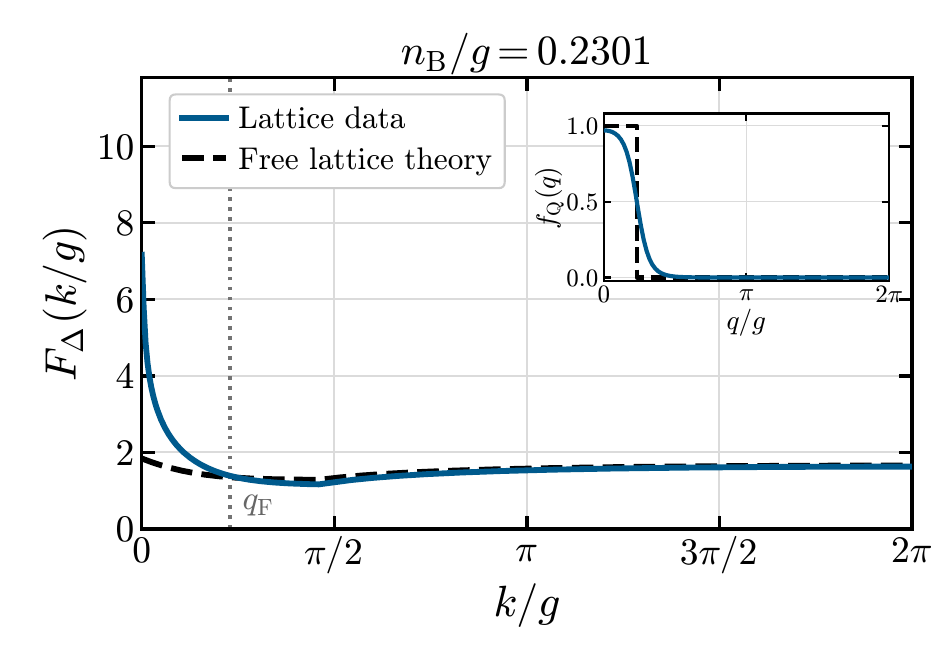}
  \caption{Diquark structure factor $F_\Delta(k/g)$ with the quark distribution $f_{\mathrm{Q}}(q)$ in the inset at $n_{\mathrm{B}}/g=0.2301$, $m/g=0.5$, and $ga=0.25$.  Blue solid and black dashed curves show interacting and free lattice results at the same density.  The dotted line marks $k=q_{\mathrm{F}}=\pi n_{\mathrm{B}}$.}
  \label{fig:quark-diquark-intermediate}
\end{figure*}

\begin{figure*}
  \centering
  \includegraphics[width=0.85\textwidth]{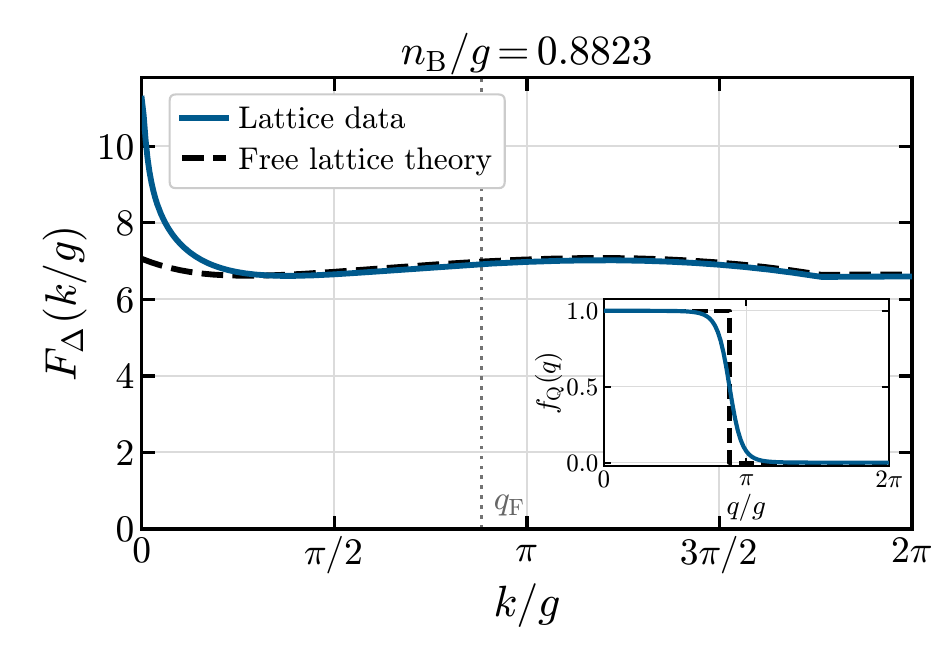}
  \caption{Same as Fig.~\ref{fig:quark-diquark-intermediate} at $n_{\mathrm{B}}/g=0.8823$.}
  \label{fig:quark-diquark-dense}
\end{figure*}

At $m/g=0.5$, the vacuum correlator decays exponentially and does not affect the large-$|n|$ behavior of $C_\Delta$.  The Luttinger parameter of the baryon-number sector is extracted from the equation of state as $K_{\mathrm{LL}} = (\pi/v_s)\,(n_{\mathrm{B}}/\muB)$~\cite{Fujikura:2026tqk}; it equals $1$ for a dilute gas of free fermions and $1/\Nc$ for free massless quarks.  The analysis in Ref.~\cite{Fujikura:2026tqk} finds that $K_{\mathrm{LL}}$ decreases from approximately $1$ toward $1/\Nc=1/2$ as the density increases.  Tomonaga--Luttinger theory then gives $K_{\mathrm{LL}}>1/2$ with the large-$|n|$ and small-$k$ behavior as
\begin{equation}
\begin{aligned}
  C_\Delta(n)
  &\propto |n|^{-1/(2K_{\mathrm{LL}})}
  && \text{for} \quad |n|\to\infty \,, \\
  F_\Delta(k/g)
  &\propto |k/g|^{1/(2K_{\mathrm{LL}})-1}
  && \text{for} \quad k\to0 \,. 
\end{aligned}
  \label{eq:diquark-tll-asymptotics}
\end{equation}
The decay exponent therefore increases from $1/2$ in the dilute baryon limit toward $1$ in the high-density quark limit, while $C_\Delta$ exhibits algebraic decay below and above the saturation density.  Figures~\ref{fig:quark-diquark-intermediate} and~\ref{fig:quark-diquark-dense} show $F_\Delta$ calculated from the equal-time lattice data, with an enhancement near $k=0$ at both densities below and above saturation.

The enhancement near $k=0$ may reflect the tendency of bosonic diquark baryons to accumulate at low total momentum.  However, Fourier orthogonality applied to Eq.~\eqref{eq:diquark_two-point_function} gives
\begin{equation}
  \int_{-\pi/(2ga)}^{\pi/(2ga)}\frac{\mathrm{d}(k/g)}{2\pi}\,F_\Delta(k/g)
  =\frac{C_\Delta(0)}{2ga\,g^2} ,
  \label{eq:diquark-structure-factor-integral}
\end{equation}
so the momentum integral is determined by the local correlator $C_\Delta(0)$ rather than the baryon density.  Although a baryon momentum distribution would help test the quarkyonic shell picture, defining it in first-principles lattice QCD is conceptually challenging.

\subsection{\texorpdfstring{QC$_2$D$_4$ at $\bm{\muB\neq0}$}{QC2D4 at muB not eq 0}}
\label{sec:2c}

We shall discuss $(3+1)$-dimensional (or Euclidean 4-dimensional) two-color QCD, for which we write QC$_2$D$_4$ here, at finite baryon density.  The bulk thermodynamics exhibits similarity to realistic nuclear matter, while baryons are not fermions but bosons.

From finite-density QC$_2$D$_4$, it has been argued that the high-density regime of this theory has properties similar to Quarkyonic Matter. The lattice studies in Refs.~\cite{Hands:2006ve, Iida:2019rah, Hands:2012fs, Boz:2019enj, Iida:2022hyy, Iida:2024irv, Braguta:2016cpw, Astrakhantsev:2018uzd,Astrakhantsev:2020tdl, Begun:2022bxj} indicate that, as the chemical potential increases, the system passes through the three regimes discussed in Sec.~\ref{sec:bec-bcs}: the hadronic (confined) phase, the BEC regime, and the BCS regime; see a recent review~\cite{Itou:2025vcy}.
Let us see how these regimes are identified from the lattice data, mainly following the recent calculation~\cite{Iida:2024irv}.  Because discussions in Ref.~\cite{Iida:2024irv} are given in terms of the quark chemical potential, $\muQ$, we use both $\muQ$ and $\muB=2\muQ$ in this subsection.

At $T=0$, the transition from the zero-density vacuum in the hadronic phase to the superfluid phase with a diquark condensate occurs at the critical baryon chemical potential,
\begin{equation}
  \mu_{\mathrm{B},\mathrm{c}} \simeq m_{\mathrm{PS}} 
  \quad \Leftrightarrow \quad
  \mu_{\mathrm{Q},\mathrm{c}} \simeq \frac{m_{\mathrm{PS}}}{2} \,.
  \label{eq:onset}
\end{equation}
Here, $m_{\mathrm{PS}}$ is the vacuum mass of the lightest pseudoscalar meson, which is degenerate with the scalar diquark carrying nonzero baryon number.
In fact, in two-color QCD, the pseudoscalar meson and the scalar diquark are degenerate at $\muB=0$, and the onset is expected to be determined by the ``pion'' mass.

Just above the onset, the system is considered to be strongly coupled, so that the BEC regime realizes.  In this density region, the order parameter of superfluidity, i.e., the diquark condensate follows the leading-order ChPT prediction~\cite{Kogut:2000ek},
\begin{equation}
  \langle qq \;\rangle\;
  \propto \left( 1-\mu_{\mathrm{Q},\mathrm{c}}^4/\muQ^4 \right)^{1/2}
  \;\sim\; \left( \frac{\muQ}{\mu_{\mathrm{Q},\mathrm{c}}} - 1 \right)^{1/2}
  \label{eq:ChPT-qq}
\end{equation}
near the onset, $\muQ\simeq \mu_{\mathrm{Q},\mathrm{c}}$.  In Ref.~\cite{Iida:2024irv}, the value of $\mu_{\mathrm{Q},\mathrm{c}}$ has been determined by treating $\mu_{\mathrm{Q},\mathrm{c}}$ as a parameter to fit the near-onset data for $\langle qq\rangle$ with this ChPT behavior.  The obtained value of $\mu_{\mathrm{Q},\mathrm{c}}/m_{\mathrm{PS}}=0.47$ is close to the expectation~\eqref{eq:onset}.

\begin{figure}
    \centering
    \includegraphics[width=0.85\textwidth]{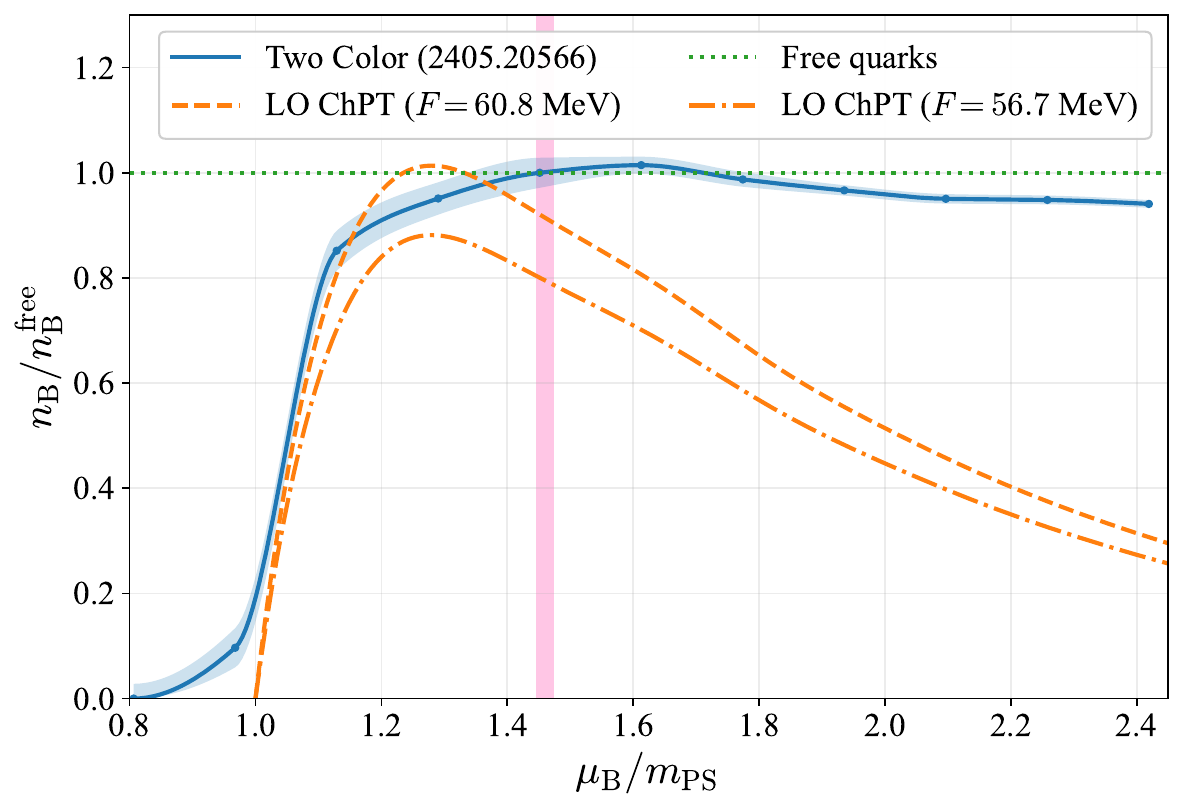}
    \caption{Dimensionless baryon density in units of the free lattice baryon density, $\nB/\nB^{\mathrm{free}}$, as a function of the dimensionless baryon chemical potential, $\muB/m_{\mathrm{PS}}$, where $\muB=2\muQ$ in the two-color case, from the data of Ref.~\cite{Iida:2024irv}.  The dotted curve is the leading-order ChPT prediction with $F = f_{\mathrm{PS}}/(2\sqrt{2}) = \qty{60.8}{\MeV}$ and $\qty{56.7}{\MeV}$ (these values are before rescaling by $0.75$; this figure itself is rescaling invariant), and the vertical pink-color band marks the BEC--BCS crossover at $\muQ/m_{\mathrm{PS}} \approx 0.73$.}
    \label{fig:nq_mu_2_3d}
\end{figure}

As discussed in Sec.~\ref{sec:bec-bcs}, the crossover from the BEC regime to the BCS regime is not a phase transition, and its location is not defined by an order parameter.  A practical criterion is provided by the quark number density; we identify the BCS regime as the region where the measured quark number density is well approximated by the free quark result in the lattice~\cite{Hands:2006ve}.  Indeed, as shown in Fig.~\ref{fig:nq_mu_2_3d}, the data of Ref.~\cite{Iida:2024irv} show that the onset chemical potential is
$\muQ/m_{\mathrm{PS}} \gtrsim 0.73$, i.e.,
\begin{equation}
  \nB / \nB^{\mathrm{free}} \simeq 1
  \quad \text{ for } \quad \muB/m_{\mathrm{PS}} \gtrsim 1.46 \,.
\end{equation}

At the onset from the BEC to the BCS regimes in this criterion, the scaling behavior of $\langle qq \rangle$ as a function of $\muQ$ also changes.  
In the BEC regime, the data are well described by the ChPT prediction mentioned above.  
In the BCS regime, the diquark condensate exhibits an approximately quadratic scaling, $\langle qq \rangle \propto \muQ^2$, as expected from weak-coupling analysis~\cite{Schafer:1999fe, Hanada:2011ju, Kanazawa:2013crb}.
The curvature of $\langle qq\rangle$ with respect to $\muQ$ in the BCS regime is completely different from that of the ChPT fit in the BEC regime.  In the BCS regime, a weak-coupling description in terms of quasi-free quarks near the Fermi surface becomes more appropriate.

Historically, the high-density region, namely the BCS regime, of two-color QCD was sometimes expected to become deconfined~\cite{Hands:2006ve, Cotter:2012mb}.  A coarse-lattice study suggested deconfinement at a quark chemical potential of approximately $\qty{0.8}{\GeV}$~\cite{Cotter:2012mb}, whereas a subsequent finer-lattice study shifted the inferred transition to higher chemical potential~\cite{Begun:2022bxj}.  
For a pion mass of $\qty{378}{\MeV}$, BCS-like scaling was observed to begin at $\muQ = 0.5\text{–}\qty{0.6}{\GeV}$, while the system retained confining behavior up to $\muQ \sim \qty{1.4}{\GeV}$~\cite{Braguta:2016cpw}.  Static quark–antiquark interactions were also investigated in a heavier-pion setup at high density~\cite{Astrakhantsev:2018uzd}.
One important remark is that, at very large chemical potential, especially around $a\muQ \sim 1$, severe lattice artifacts should properly be taken into account.  Since the lattice spacing $a$ provides the UV cutoff, the dimensionless parameter $a\muQ$ should be kept smaller than unity for reliable continuum physics.  In several simulations, the diquark condensate, which is expected to increase with $\muQ$, was observed to decrease around $a\muQ \sim 1$~\cite{Kogut:2001na, Hands:2006ve, Iida:2019rah}. 
This behavior is now understood as a lattice artifact. The artifact can also affect the Polyakov loop, which may increase at large $a\mu$ even if this does not represent physical deconfinement.  Indeed, larger-volume and finer-lattice calculations have shown that the region where confinement persists extends to higher chemical potential than was suggested by earlier studies~\cite{Boz:2019enj, Begun:2022bxj}.

Now, the static quark-antiquark potential continues to exhibit confining behavior at long distances~\cite{Ishiguro:2021yxr,Begun:2022bxj}.  In addition, the topological susceptibility measured from the gluonic definition of the topological charge does not show strong suppression which is typical in the high-temperature deconfined phase.  Instead, even in the BCS regime, the susceptibility stays close to the value observed in the ordinary QCD vacuum at $T=\muB=0$~\cite{Iida:2019rah,Iida:2024irv}.  These results suggest that the high-density BCS regime is not simply a weakly coupled deconfined quark gas. Rather, it realizes the similarity to Quarkyonic Matter: quark degrees of freedom govern the bulk thermodynamics, while the pairing near the Fermi surface occurs in the color-singlet diquark channel and the gluonic sector remains confining and nonperturbative.

In Refs.~\cite{Iida:2024irv, Itou:2025jqx}, computations for smaller temperature and better resolution have been performed. 
For QC$_2$D$_4$ in Ref.~\cite{Iida:2024irv}, the physical scale was fixed by taking the chiral transition temperature at $T=\muB=0$ to be $T_{\mathrm{c}} = \qty{200}{\MeV}$~\cite{Iida:2020emi}.  In the analysis in this work, in contrast, we employ $T_{\mathrm{c}} = \qty{150}{\MeV}$, closer to the QCD crossover temperature, which 
leads to a rescaling factor, $0.75$ as mentioned in Sec.~\ref{sec:intro}, and corresponds to a lattice spacing of $a = \qty{0.22}{\femto\meter}$.  This changes the pseudoscalar mass to $m_{\mathrm{PS}} \approx \qty{550}{\MeV}$.  With this setting of scale, the density can be computed as a function of the chemical potential and compared to that of a free gas of quarks.
We see from Fig.~\ref{fig:nq_mu_2_3d} good agreement with the density of a free gas of quarks when
$\muB/m_{\mathrm{PS}} \ge 1.2$, corresponding to a quark chemical potential of $\muQ \sim \qty{300}{\MeV}$.

It should be noted that this good agreement is not a robust feature.  In Fig.~\ref{fig:nq_mu_2_3d}, the ChPT predictions (with two different scale settings) are depicted by the orange lines, and their peaks are located around $\muB/m_{\mathrm{PS}} \approx 1.3$ with the peak height $\nB/\nB^{\text{free}}=0.9$-$1$.  It is possible to express the peak height and location analytically as
\begin{equation}
  \frac{\nB}{\nB^\text{free}} \biggr|_{\text{ChPT peak}} = \frac{32\pi^2 F^2}{\sqrt{3} m_{\mathrm{PS}}^2} \,,
  \qquad
  \frac{\muB}{m_{\mathrm{PS}}} \biggr|_{\text{ChPT peak}} = 3^{1/4} \approx 1.32 \,.
  \label{eq:ChPT_peak}
\end{equation}
For the setup of Ref.~\cite{Iida:2024irv}, $F\approx \qty{60}{\MeV}$ and $m_{\mathrm{PS}}\approx \qty{740}{\MeV}$ (before applying the rescaling factor, 0.75), the peak height estimated from the above expression happens to be close to unity.\footnote{Strictly speaking, $\nB^{\text{free}}$ in Fig.~\ref{fig:nq_mu_2_3d} is estimated on the $32^4$ lattice, while Eq.~\eqref{eq:ChPT_peak} adopts the continuum expression, which makes quantitative differences by $15$-$20\,\%$.}
However, the peak height is sensitive to $m_{\mathrm{PS}}$, and if it is close to the physical pion mass, the peak must have been far more enhanced, and $\nB/\nB^{\text{free}} \sim 1$ would occur at much higher density, as is actually the case in the lattice-QCD results at high isospin density; see Fig.~\ref{fig:ni_mui}.  Nevertheless, this lattice setup with heavy $m_{\mathrm{PS}}$ turns out to be convenient for our purpose of detecting the BEC-BCS crossover.
In particular, we would stress that the departure from the ChPT is more prominent than seen in Fig.~\ref{fig:vs_comparison}.  
In fact, the condition of $\nB/\nB^\text{free} \sim 1$ by itself is not crucial for our understanding of Quarkyonic Matter.  It is a reasonable interpretation that the ChPT breakdown for $\muB/m_{\mathrm{PS}} \gtrsim 1.32$ signifies the onset of quark degrees of freedom, possibly the indication of Quarkyonic Matter.   
In QC$_2$D$_4$, the quarkyonic onset happens to be located near the density where $\nB/\nB^\text{free} \sim 1$, but this is not necessarily the case, as is evident in the high-isospin data.

The sound velocity, obtained from the same set of lattice data, was already shown in Fig.~\ref{fig:vs_comparison} with the density rescaled by $2\alpha/3$.  The original analysis was performed in Ref.~\cite{Iida:2024irv}.  In the relatively low-density BEC regime, the data again agree with the ChPT prediction.  This agreement is therefore a nontrivial check that the near-onset region is well described by the BEC picture.
In the BCS regime, on the other hand, the lattice data clearly deviate from the ChPT prediction.  This discrepancy is visible, in particular for the $T=\qty{40}{\MeV}$ data (the two-color curve in Fig.~\ref{fig:vs_comparison} is based on the $T=\qty{40}{\MeV}$ data, which is scaled down to $T=\qty{30}{\MeV}$ with the rescaling factor, 0.75), suggesting that the high-density superfluid regime is qualitatively different from the near-onset BEC regime.  Around the BEC--BCS crossover, $\muQ/m_{\mathrm{PS}}\simeq 0.73$, the sound velocity squared already exceeds $1/3$, as seen from the original plot in Fig.~11 of Ref.~\cite{Iida:2024irv}.  This behavior may well be interpreted as the onset of Quarkyonic Matter, as argued above.

\subsection{\texorpdfstring{QC$_3$D$_4$ at $\bm{\muI\neq0}$}{QC3D4 at muI not eq 0}}

\begin{figure}
    \centering
    \includegraphics[width=0.9\textwidth]{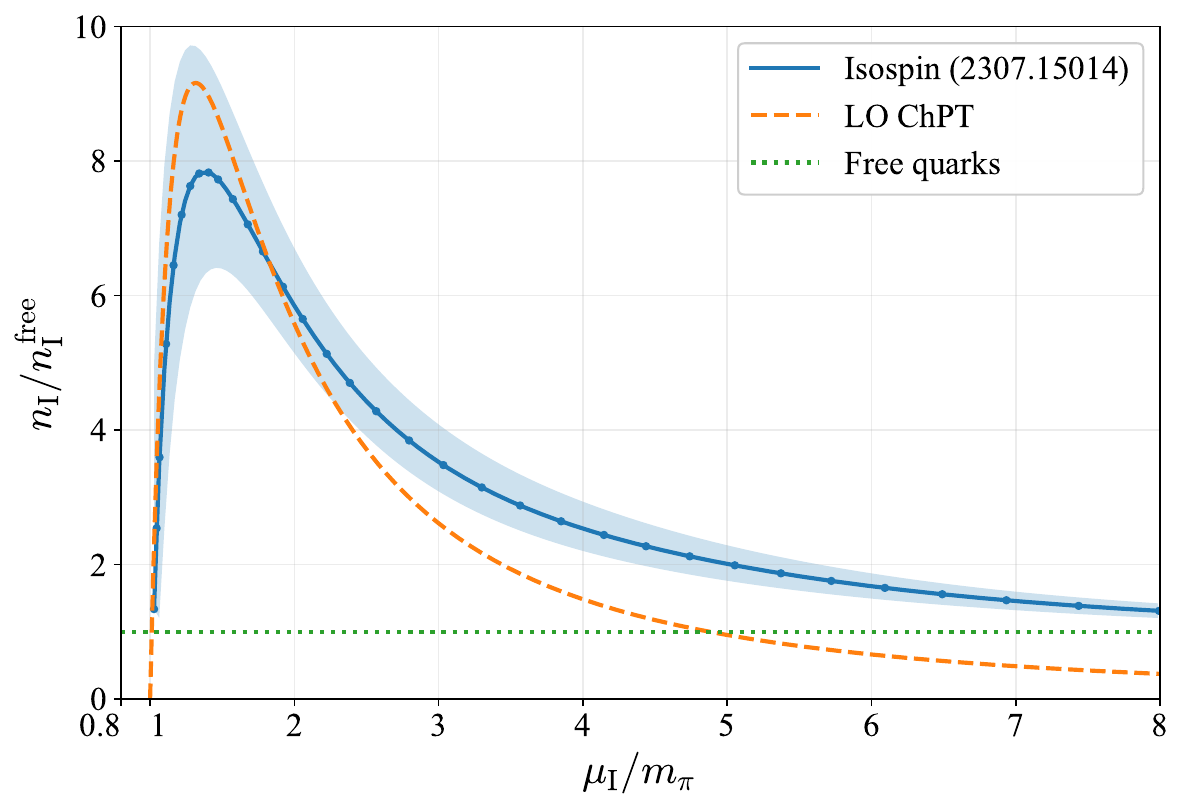}
    \caption{The ratio of the isospin density to the free-quark value as a function of the isospin chemical potential, based on the lattice results of Refs.~\cite{Abbott:2023coj,Abbott:2024vhj}.  The leading-order ChPT prediction and the free-quark value are shown for comparison.}
    \label{fig:ni_mui}
\end{figure}

We now turn to three-color QCD at finite isospin chemical potential.  The excess of $u$ quarks and $\bar d$ antiquarks forms a charged-pion condensate.  Lattice calculations provide the isospin density and EoS over a wide range of chemical potentials~\cite{Abbott:2023coj,Abbott:2024vhj}.  Near the onset of pion condensation, the results agree with leading-order ChPT formulas.  For the density region shown in Fig.~\ref{fig:ni_mui}, ChPT predicts a maximum at
\begin{equation}
    \left.\frac{\nI}{\nI^{\text{free}}}\right|_{\text{ChPT peak}}
    = \frac{8\pi^2 f_\pi^2}{3\sqrt{3}m_\pi^2} \, ,
    \qquad
    \left.\frac{\muI}{m_\pi}\right|_{\text{ChPT peak}}
    = 3^{1/4} \approx 1.32 \,,
\end{equation}
which is the isospin counterpart of Eq.~\eqref{eq:ChPT_peak}.
The lattice density begins to depart appreciably from the ChPT curve above $\muI/m_\pi\sim 2$.  Thus, in this case, the ChPT maximum and the departure of the lattice data from ChPT are distinct features.

As in the two-color case discussed in Sec.~\ref{sec:2c}, we interpret the departure from ChPT as a possible signature of the onset of quark degrees of freedom, associated with Quarkyonic Matter.  In the picture developed in Sec.~\ref{sec:background}, the pion condensate describes the low-density system, while increasing isospin density fills the available quark states.  The saturation criterion for the quark distribution, Eq.~\eqref{eq:KFC_isospin}, then requires some change in $f_{\mathrm{M}}(\bk)$.  The observed breakdown of the pion-based ChPT description is consistent with the expected change with the growing importance of quark degrees of freedom.  It does not, by itself, establish that the quark occupation has reached the saturation criterion, though.

Figure~\ref{fig:ni_mui} also shows why the onset of this regime should not be identified with the onset of an ideal quark gas.  Although the lattice density bends toward the free-quark result, it remains substantially above it: $\nI/\nI^{\text{free}}$ is about $2.4$ at $\muI/m_\pi = 4$ and about $1.4$ at $\muI/m_\pi = 7.5$, as seen in Fig.~\ref{fig:ni_mui}.  The data therefore allow for a broad intermediate region after the departure from ChPT and before the density approaches the quasi-free-quark value.  This region is wider than the corresponding window seen in the two-color setup of Sec.~\ref{sec:2c}.

The sound velocity provides a complementary indication of the change in behavior.  It reaches a peak around $\muI/m_\pi = 2\text{-}2.5$, close to where the density begins to deviate from ChPT~\cite{Abbott:2023coj}.  At higher chemical potential, $v_s^2$ decreases but remains above the conformal value $1/3$ over a wide range, extending to approximately $\muI/m_\pi\sim 14$ in Ref.~\cite{Abbott:2023coj}.  We interpret these observations together as indirect evidence for the onset of a quarkyonic-saturation regime.
Whether this high-isospin-density regime realizes Quarkyonic Matter remains to be established directly, though a theoretical argument supports confinement.  According to Refs.~\cite{Son:2000xc,Son:2000by,Cohen:2015soa}, at asymptotically high isospin density and zero temperature, quark excitations are gapped by the color-singlet quark-antiquark pairing.  The physical system below this gap scale is effectively described by pure gluonic theory, which remains confining, with a confinement scale much smaller than the gap scale (for a quantitative estimate, see Ref.~\cite{Cohen:2015soa}).  This provides a theoretical basis for expecting confinement at high isospin density. Direct measurements of the quark distribution and confinement-sensitive observables would provide further tests of the quarkyonic saturation interpretation.

\section{Summary and Conclusions}

We have examined whether dense QCD-like matter exhibits the characteristic separation of degrees of freedom expected in a quarkyonic picture: quarks increasingly govern bulk thermodynamics, while color-singlet excitations and nonperturbative gauge dynamics remain important.  In the distribution-function model reviewed in Sec.~\ref{sec:background}, the central organizing idea is quarkyonic saturation.  Once the occupation of low-momentum quark states reaches its bound due to Pauli blocking, additional density changes the distribution of the composite degrees of freedom and builds up a quark-dominated bulk.  A stiff EoS and a peak in $v_s^2$ can accompany this rearrangement.  These are predictions of the picture; thermodynamic data alone do not establish that the quark occupation has saturated.

The $(1+1)$-dimensional two-color theory offers the most direct test considered here.  Its lattice quark distribution approaches unit occupation at low momentum around $\nB/g\approx 0.2$, while the bulk density crosses the free-lattice-quark curve only at higher density, around $\nB/g\approx 0.38$; see Sec.~\ref{sec:lattice}.  For massless quarks, the equal-time distribution has a convolution form with a density-independent kernel.  A one-parameter kernel model describes the lattice distributions, and its functional form also fits the massive data when the scale is allowed to vary with density.  The diquark structure factor retains an enhancement at low momentum on both sides of the saturation region.  Its integral, however, is not a baryon number sum rule, so these correlations do not yet determine a baryon momentum distribution or demonstrate the shell structure of the quarkyonic model.

In $(3+1)$-dimensional two-color QCD, the diquark-condensed state agrees with leading-order ChPT close to onset and departs from it at higher chemical potential.  The density and condensate then acquire BCS-like behavior, and the density approaches the free-quark result~\cite{Iida:2024irv}.  At the same time, confinement-sensitive observables in the available lattice studies retain substantial nonperturbative behavior~\cite{Braguta:2016cpw,Boz:2019enj,Begun:2022bxj,Iida:2024irv}.  This coexistence is consistent with the quarkyonic interpretation.  The location of its onset remains lattice parameter dependent, and quantitative comparisons must account for the relatively heavy pseudoscalar mass, finite temperature (physical scale setting), and lattice-spacing effects.

For three-color QCD at finite isospin density, the distinction between the breakdown of ChPT and the approach to a free quark gas is particularly clear.  The lattice density begins to depart from ChPT around $\muI/m_\pi\sim 2$, near the peak in the sound velocity, but $\nI/\nI^{\text{free}}$ remains above unity throughout the range displayed in Fig.~\ref{fig:ni_mui}~\cite{Abbott:2023coj}.  The sound velocity subsequently falls only gradually toward the conformal value.  We interpret the density region for $\muI/m_\pi \gtrsim 2$ as a candidate quarkyonic-saturation regime, in which quark degrees of freedom become increasingly relevant before the thermodynamics resembles a weakly interacting quark gas.  Confinement and saturation have not yet been measured directly in this regime.

The sound velocities inferred for neutron-star matter and calculated in the two lattice systems become comparably large over part of the density range shown in Fig.~\ref{fig:vs_comparison}~\cite{Fujimoto:2024cyv,Iida:2024irv,Abbott:2023coj}.  This comparison is suggestive, but the horizontal axis for the isospin system and the rescaled two-color system is an equivalent baryon density introduced for comparison, not the physical baryon density of neutron-star matter.  More importantly, baryons in these lattice systems are bosons, though nuclear matter is composed of fermionic degrees of freedom.  Their onset densities therefore need not coincide.  The similarity of the high-density sound velocities is a motivation to investigate a common saturation mechanism, rather than a determination of one from the EoS alone.

The decisive next steps are to calculate baryon and gauge-invariant quark distributions in the $(3+1)$-dimensional systems and to establish whether low-momentum occupation exhibits expected behavior at saturation where the lattice results depart from ChPT\@.  Measurements sensitive to confinement at finite isospin density are also needed.  Baryon and pion distribution functions, or suitably defined correlation observables that constrain them, would test the predicted redistribution of color-singlet degrees of freedom.  Together with continuum and lighter-mass studies, these measurements would distinguish the proposed quarkyonic interpretation from other mechanisms that can produce similar thermodynamic behavior.

\acknowledgments
The authors thank
Toru~Kojo and
Shuhei~Minato
for useful discussions.
They also thank the Yukawa Institute for Theoretical Physics at Kyoto University for its hospitality, where this work was initiated during the YITP-W-26-04 on ``Buenas Ideas on the QCD Phase Diagram.''
Larry~McLerran acknowledges partial support from the Institute for Nuclear Theory, which is funded in part by the U.S.\ Department of Energy grant No.\ DE-FG02-00ER41132.
The work of Kohei~Fujikura, Yoshimasa~Hidaka, and Etsuko~Itou was supported by JST, CREST Grant Number JPMJCR24I3.
This work is also supported by JSPS KAKENHI, Grant Numbers JP26K17144 (K.~Fujikura), JP22H05118 (K.~Fukushima), JP25K24464 (K.~Fukushima, Y.~H.), JP26K00698 (K.~Fukushima), JP24H00975 (Y.~H.), JP25K01002 (Y.~H.), JP23H05439 (E.~I.)
and JP25K01001(E.~I.),
JST Grant Numbers JPMJPF2221 (E.~I.).
The work of E.~I. is supported by 
the Program for Promoting Researches on the Supercomputer ``Fugaku'' (Simulation for basic science: from fundamental laws of particles to creation of nuclei, and Simulation for basic science: approaching the new quantum era), and the Joint Institute for Computational Fundamental Science (JICFuS), Grant Number JPMXP1020230411. 
This work is also supported by the Center for Gravitational Physics and Quantum Information (CGPQI) at Yukawa Institute for Theoretical Physics.
The authors used Claude Code (Anthropic, Claude Fable 5, Claude Opus 5, Claude Opus 5.5, and Claude Sonnet 5), ChatGPT and Codex (OpenAI, gpt-5.6-sol, gpt-6-sol, and gpt-6-astra), and Gemini (Google, Gemini 4 Argon) in preparing this paper.  These tools were used for language editing, revising the text, checking derivations and numerical estimates against independent symbolic and numerical calculations, developing and revising code for numerical calculations and data analysis, compiling the data for Figs.~\ref{fig:vs_comparison}, \ref{fig:muBnB}, \ref{fig:nq_mu_2_3d}, \ref{fig:ni_mui}, preparing figures including Figs.~\ref{fig:qdist_fitform}, \ref{fig:quark-diquark-intermediate}, and \ref{fig:quark-diquark-dense}, and checking the bibliography.  Checks of the numerical code included free theory benchmarks and comparisons with independent direct calculations of operator matrix elements and correlators on short lattice segments.  The authors directed the use of these tools, reviewed and revised the manuscript, and take full responsibility for its content.

\appendix

\section{Quark distribution function in the massless two-flavor Schwinger model}
\label{app:schwinger2f}

We use the exact bosonized solution of the massless two-flavor Schwinger model to compute its gauge-invariant quark distribution function at finite isospin chemical potential. See, e.g., Refs.~\cite{Schwinger:1962tp,Coleman:1975pw,GKMS} for the Schwinger models with a bosonization technique in the absence of the isospin chemical potential.

\subsection{The model and the observable}
\label{app:model}

The model contains a $\mathrm{U}(1)$ gauge field $A_\mu$ with $F_{\mu\nu}\coloneqq\partial_\mu A_\nu-\partial_\nu A_\mu$ and $\Nf=2$ massless unit-charge Dirac fermions $\psi_u,\psi_d$ at isospin chemical potential $\muI$.  The gauge coupling $g_{\mathrm{U}(1)}$ sets the intrinsic scale.  We use the metric $\mathrm{diag}(+1,-1)$, $\epsilon^{01}=+1$, and $\gamma^0=\sigma^1$, $\gamma^1=-\ri\sigma^2$.  The Lagrangian is
\begin{equation}
  \mathcal{L}
  = -\frac{1}{4}F_{\mu\nu}F^{\mu\nu}
  + \sum_{f} \bar\psi_f\,\ri\gamma^\mu\big(\partial_\mu + \ri g_{\mathrm{U}(1)} A_\mu\big)\psi_f
  + \muI\, I_3 \,,
  \qquad
  I_3 = \frac{1}{2}\big(\psi_u^\dagger\psi_u - \psi_d^\dagger\psi_d\big) \,.
  \label{eq:app:model}
\end{equation}
The isospin density $I_3$ is the baryon-density analogue in this model.  The product $\gamma^5=\gamma^0\gamma^1=\sigma^3$ has eigenvalues $s=\pm$ that label chirality, and $\psi_{f,s}$ denotes the component of $\psi_f$ with $\gamma^5=s$.  The flavor chemical potentials are $\mu_u=+\muI/2$ and $\mu_d=-\muI/2$.

We define the equal-time correlator $W(x)$ using the Wilson line $U(x):=e^{-\ri g_{\mathrm{U}(1)}\int_0^x A_1\, dz}$ as
\begin{align}
  W(x) := \frac{1}{2}\Big[
    \big\langle \psi_u^\dagger(x)\,U(x)\,\psi_u(0)\big\rangle
   -\big\langle \psi_d^\dagger(x)\,U(x)\,\psi_d(0)\big\rangle \Big] \,.
\end{align}
Then, we also define the quark distribution as follows:
\begin{align}
  f_{\mathrm{Q}}(q) := \int dx\; e^{-\ri q x}\, W(x) \,.
  \label{eq:app:def}
\end{align}
The fermion fields are evaluated at equal time with contracted spinor indices.  The expectation value is taken in the ground state of Eq.~\eqref{eq:app:model} at nonzero $\muI$.  The Wilson line makes each correlator gauge invariant.

The equality $W(0)=\langle I_3\rangle$ identifies $W(0)$ with the isospin density and fixes the normalization of $f_{\mathrm{Q}}(q)$.  The $d$-flavor correlator follows from the $u$-flavor correlator by $\muI\to-\muI$.

\subsection{Abelian bosonization}
\label{app:boson}

Bosonization maps each Dirac fermion to a compact real scalar $\phi_f$ of period $\sqrt{\pi}$.  Let $\Pi_f$ denote the momentum conjugate to $\phi_f$.  The dual field $\theta_f$ is defined at equal time by
\begin{equation}
  \partial_1\theta_f \coloneqq -\Pi_f \,.
  \label{eq:app:dual-spatial}
\end{equation}
For a free massless scalar, we choose the additive zero mode such that
\begin{equation}
  \partial_0\theta_f = -\partial_1\phi_f \,.
  \label{eq:app:dual-temporal}
\end{equation}
The interchange of temporal and spatial derivatives motivates the term dual field.
With point-splitting cutoff $a$ and Klein factors $\eta_{f,s}$, the bosonization dictionary is
\begin{align}
  \psi_{f,s}~ &\leftrightarrow ~ \frac{\eta_{f,s}}{\sqrt{2\pi a}}\,
  e^{\,\ri\sqrt{\pi}\left[s\,\phi_f+\theta_f\right]} \,,
  \label{eq:app:dictvertex}\\
  {:}\bar\psi_f\gamma^\mu\psi_f{:}~ &\leftrightarrow ~ \frac{1}{\sqrt{\pi}}\,\epsilon^{\mu\nu}\partial_\nu\phi_f \,.
  \label{eq:app:dictcurrent}
\end{align}
The Klein factors enforce anticommutation between distinct species.  We use the correspondence of Refs.~\cite{Coleman:1974bu,Mandelstam:1975hb} and the gauged multiflavor solution of Ref.~\cite{HetrickHosotaniIso}.

For $X\in\{\phi,\theta,\Pi\}$, define $X_{c,I}\coloneqq(X_u\pm X_d)/\sqrt{2}$.  This orthogonal rotation preserves $\partial_1\theta_{c,I}=-\Pi_{c,I}$ and diagonalizes the gauge and chemical-potential couplings.  Equation~\eqref{eq:app:dictcurrent} then gives $I_3=\partial_1\phi_I/\sqrt{2\pi}$.
The bosonized Lagrangian is given by
\begin{equation}
  \mathcal{L}
  = -\frac{1}{4}F_{\mu\nu}F^{\mu\nu}
  + \frac{1}{2}\big(\partial_\mu\phi_c\big)^2
  + \frac{1}{2}\big(\partial_\mu\phi_I\big)^2
  + g_{\mathrm{U}(1)}\sqrt{\frac{\Nf}{\pi}}\,\epsilon^{\mu\nu}\partial_\mu\phi_c\,A_\nu
  + \frac{\muI}{\sqrt{2\pi}}\,\partial_1\phi_I \,.
  \label{eq:app:Lbos}
\end{equation}

The momenta conjugate to $\phi_c$, $\phi_I$, and $A_1$ are
\begin{equation}
  \Pi_c = \partial_0\phi_c + g_{\mathrm{U}(1)}\sqrt{\frac{\Nf}{\pi}}\, A_1 \,,
  \qquad
  \Pi_I = \partial_0\phi_I \,,
  \qquad
  \Pi_A = E \coloneqq F_{01} \,.
  \label{eq:app:momenta}
\end{equation}
Varying the Lagrange multiplier $A_0$ gives Gauss's law constraint,
\begin{equation}
  \partial_1 E = g_{\mathrm{U}(1)}\sqrt{\frac{\Nf}{\pi}}\, \partial_1\phi_c \,.
  \label{eq:app:gausslaw}
\end{equation}
We impose this constraint on the physical states.
With $E(\infty)=0$ and $\phi_c(\infty)=0$, Eq.~\eqref{eq:app:gausslaw} gives
\begin{equation}
  E = g_{\mathrm{U}(1)}\sqrt{\frac{\Nf}{\pi}}\, \phi_c \,.
  \label{eq:app:gauss}
\end{equation}
This eliminates the electric field from the Hamiltonian.  We choose the axial gauge $A_1=0$, where $\Pi_c=\partial_0\phi_c$ and $E=-\partial_1 A_0$.
Consequently, the gauge boson acquires the following mass,
\begin{equation}
  m_\gamma \coloneqq g_{\mathrm{U}(1)}\sqrt{\frac{\Nf}{\pi}}  \quad (\Nf=2) \,.
  \label{eq:app:schwinger-mass}
\end{equation}
The resultant Hamiltonian is now expressed as
\begin{equation}
  H = \int dx\;\left[\frac{1}{2}\Pi_c^2 + \frac{1}{2}\big(\partial_1\phi_c\big)^2 + \frac{1}{2}\,m_\gamma^2\,\phi_c^2 + \frac{1}{2}\Pi_I^2 + \frac{1}{2}\big(\partial_1\phi_I\big)^2 - \frac{\muI}{\sqrt{2\pi}}\,\partial_1\phi_I\right] \,.
  \label{eq:app:H}
\end{equation}
Minimizing Eq.~\eqref{eq:app:H} fixes the isospin density and the Fermi momentum,
\begin{equation}
  n_I \coloneqq \langle I_3\rangle = \frac{\langle\partial_1\phi_I\rangle}{\sqrt{2\pi}} = \frac{\muI}{2\pi} ,
  \qquad
  q_{\mathrm{F}} \coloneqq \pi n_I = \frac{\muI}{2} \,.
  \label{eq:app:kf}
\end{equation}
This theory is Gaussian, so it is exactly solvable.

\subsection{Quark distribution function}
\label{app:closed}

The bosonization mapping introduced in the previous subsection enables us to compute the fermion correlators in Eq.~\eqref{eq:app:def}.  With $\lambda\coloneqq\sqrt{\pi/\Nf}$, the mapping for the $u$ field is
\begin{equation}
  \psi_{u,s}(x) ~ \leftrightarrow ~ \frac{\eta_{u,s}}{\sqrt{2\pi a}}\;
  e^{\,\ri\lambda\left[s\,\phi_c(x)+\theta_c(x)\right]}\,
  e^{\,\ri\lambda\left[s\,\phi_I(x)+\theta_I(x)\right]} \,.
  \label{eq:app:vertexu}
\end{equation}
In the axial gauge $A_1=0$, the Wilson line is $U(x)=1$.
Equation~\eqref{eq:app:vertexu} gives
\begin{equation}
\begin{aligned}
  C_{\alpha,s}^{(a)}(x)
  &\coloneqq \Big\langle
  e^{-\ri\lambda\left[s\phi_\alpha(x)+\theta_\alpha(x)\right]}
  e^{\,\ri\lambda\left[s\phi_\alpha(0)+\theta_\alpha(0)\right]}
  \Big\rangle \,, \\
  G_{u,s}(x)
  &\coloneqq \big\langle\psi_{u,s}^\dagger(x)\psi_{u,s}(0)\big\rangle
  = \lim_{a\to0}\frac{1}{2\pi a}
  C_{c,s}^{(a)}(x)C_{I,s}^{(a)}(x)
\end{aligned}
  \label{eq:app:Wbos}
\end{equation}
for $x\ne0$.
Here, $\alpha=c,I$.
Expectation values $\langle \cdots \rangle$ are taken at equal time in the ground state of Eq.~\eqref{eq:app:H}.
The point-splitting cutoff $a$ in Eq.~\eqref{eq:app:dictvertex} is implemented by the factor $e^{-|\ell|a}$ in integrals over the bosonic momentum $\ell$.

Let $\langle\,\cdots\,\rangle_m$ denote the ground-state expectation value of a free scalar field of mass $m$. Using $\partial_1\theta_c=-\Pi_c$ and the Baker-Campbell-Hausdorff identity for the charge-vertex commutator gives
\begin{equation}
\begin{aligned}
  \mathcal{X}_s(x)
  &\coloneqq \lambda\left[s\big(\phi_c(x)-\phi_c(0)\big)
  +\theta_c(x)-\theta_c(0)\right]  , \\
  \mathcal{D}(m_\gamma|x|)
  &\coloneqq \lim_{a\to0}
  \frac{\big\langle e^{-\ri\mathcal{X}_s(x)}\big\rangle_{m=m_\gamma}}
  {\big\langle e^{-\ri\mathcal{X}_s(x)}\big\rangle_{m=0}}
  = \exp\left\{-\frac{1}{2}\lim_{a\to0}\left[
  \big\langle\mathcal{X}_s^2(x)\big\rangle_{m=m_\gamma}
  -\big\langle\mathcal{X}_s^2(x)\big\rangle_{m=0}
  \right]\right\}  \,.
\end{aligned}
  \label{eq:app:Dratio}
\end{equation}
The zero-temperature integral evaluated in Ref.~\cite{SVZ} gives
\begin{equation}
  \lim_{a\to0}\left[
  \big\langle\mathcal{X}_s^2(x)\big\rangle_{m=m_\gamma}
  -\big\langle\mathcal{X}_s^2(x)\big\rangle_{m=0}\right]
  = \frac{\lambda^2}{\pi}\big[-2B(m_\gamma|x|)+\Xi(m_\gamma|x|)\big] \,,
  \label{eq:app:vertex-variance}
\end{equation}
\begin{equation}
  B(u) \coloneqq K_0(u)+\ln\frac{u}{2}+\gamma_E \,, \qquad
  \Xi(u) \coloneqq \int_0^u dt\;(u-t)K_0(t) \,.
  \label{eq:app:B-Xi}
\end{equation}
Here, $\gamma_E$ is Euler's constant and $K_0$ is the modified Bessel function of the second kind.  Substitution gives
\begin{equation}
  \mathcal{D}(u)
  = \exp\Big\{\kappa\big[2B(u)-\Xi(u)\big]\Big\} \,, \qquad
  \kappa \coloneqq \frac{\lambda^2}{2\pi}=\frac{1}{2\Nf} \,.
  \label{eq:app:D}
\end{equation}
Equations~\eqref{eq:app:Wbos} and \eqref{eq:app:kf}, together with the free massless correlator of Refs.~\cite{Coleman:1974bu,Mandelstam:1975hb}, give
\begin{equation}
\begin{aligned}
  G_{u,s}(x)
  &= \frac{\ri s}{2\pi x}\,e^{-\ri sq_{\mathrm{F}}x}\mathcal{D}(m_\gamma|x|) \,, ~
  G_{d,s}(x)
  &= \frac{\ri s}{2\pi x}\,e^{\,\ri sq_{\mathrm{F}}x}\mathcal{D}(m_\gamma|x|) \,.
\end{aligned}
  \label{eq:app:flavor-correlators}
\end{equation}
Substitution into Eq.~\eqref{eq:app:def} yields
\begin{equation}
\begin{aligned}
  W(x)
  = \frac{1}{2}\sum_{s=\pm}\big[G_{u,s}(x)-G_{d,s}(x)\big]
  = \frac{\sin(q_{\mathrm{F}}x)}{\pi x}\,\mathcal{D}(m_\gamma|x|) \,,
  \qquad x\ne0 \,.
\end{aligned}
  \label{eq:app:main}
\end{equation}
The above expression is the central result of this appendix.

The small-$u$ limit of Eq.~\eqref{eq:app:D} and the point-splitting prescription in Eq.~\eqref{eq:app:Wbos} give
\begin{equation}
  \lim_{u\to0^+}\mathcal{D}(u)=1 \,, \qquad
  W(0)=\lim_{x\to0}W(x)=\frac{q_{\mathrm{F}}}{\pi}=n_I \,.
  \label{eq:app:short-distance}
\end{equation}
Without the gauge interaction ($g_{\mathrm{U}(1)}=0$), we recover the standard free-fermion result at finite density,
\begin{equation}
  W_{\mathrm{free}}(x) = \frac{\sin q_{\mathrm{F}} x}{\pi x} \,,
  \qquad
  f_{\mathrm{Q}}^{\mathrm{free}}(q) = \Theta(q_{\mathrm{F}} - |q|) \,, \qquad
  \Theta(0) \coloneqq \frac{1}{2} \,.
  \label{eq:app:free-distribution}
\end{equation}

Fourier transform of Eq.~\eqref{eq:app:main} gives the convolution of the free Fermi distribution with a density-independent kernel,
\begin{equation}
\begin{aligned}
  f_{\mathrm{Q}}(q)
  &= \int \frac{dq'}{2\pi}\; f_{\mathrm{Q}}^{\mathrm{free}}(q')K_{\text{U(1)}}(q-q') \,,~
  K_{\text{U(1)}}(q)
  \coloneqq \int dx\; e^{-\ri qx}\, \mathcal{D}(m_\gamma|x|)\,.
\end{aligned}
  \label{eq:app:conv}
\end{equation}
For numerical evaluation, the evenness of $\mathcal{D}(m_\gamma|x|)$ gives
\begin{equation}
\begin{aligned}
  f_{\mathrm{Q}}(q)
  &= S\Big(\frac{q_{\mathrm{F}}+q}{m_\gamma}\Big)
   + S\Big(\frac{q_{\mathrm{F}}-q}{m_\gamma}\Big)\,,~S(w)
  \coloneqq \frac{1}{\pi}\int_0^\infty du\;
  \frac{\sin(wu)}{u}\,\mathcal{D}(u) \,.
\end{aligned}
  \label{eq:app:np}
\end{equation}
Together with Eq.~\eqref{eq:app:D}, this gives an explicit integral representation of the quark distribution in the massless two-flavor Schwinger model.

\section{Massless quarks in (1+1)-dimensional \texorpdfstring{$\mathrm{SU}(\Nc)$}{SU(Nc)} gauge theory at finite baryon chemical potential}
\label{app:factorization}

Motivated by the fact that the Schwinger model at finite isospin chemical potential admits an exact expression for the quark distribution function, we expect that the quark distribution function of $(1+1)$-dimensional QCD with massless fundamental quarks at finite baryon chemical potential has a similar property.
Indeed, in this appendix, we show that the convolution form of Eq.~\eqref{eq:main:conv} also holds for massless continuum $\mathrm{SU}(\Nc)$ quarks.
The restriction to one spatial dimension plays a crucial role in the following derivation.

Consider one massless fundamental Dirac flavor in the continuum $\mathrm{SU}(\Nc)$ gauge theory in $(1+1)$ dimensions, at zero temperature and quark chemical potential $\muQ=\muB/\Nc>0$.  We regulate space by a circle of circumference $L$ with antiperiodic fermions.  With $A_\nu\coloneqq A_\nu^aT^a$, the Lagrangian is
\begin{equation}
  \mathcal L_{\muQ}
  =-\frac{1}{4}F_{\mu\nu}^aF^{a\mu\nu}
  +\bar\psi\big(\ri\gamma^\nu D_\nu+\muQ\gamma^0\big)\psi \,,
  \qquad D_\nu=\partial_\nu+\ri gA_\nu \,,
  \label{eq:app:sunmodel}
\end{equation}
where $T^a$ are the fundamental color generators, normalized by $\operatorname{tr}(T^aT^b)=\delta^{ab}/2$, and $[D_\mu,D_\nu]=\ri gF_{\mu\nu}^aT^a$ defines the field strength.  We use the gamma matrices of Appendix~\ref{app:model} and denote the components of $\psi$ with eigenvalues $s=\pm1$ under $\gamma^5=\gamma^0\gamma^1$ by $\psi_s$.  With $E^a\coloneqq F_{01}^a$, the Hamiltonian is
\begin{equation}
  H_{\muQ}[\psi,A]
  =\int_0^L dx\left[
  \frac{1}{2}E^aE^a-A_0^a(D_1E)^a
  +\psi^\dagger\big(-\ri\gamma^5D_1+gA_0-\muQ\big)\psi\right] \,.
  \label{eq:app:canonicalhamiltonian}
\end{equation}
Here, the adjoint covariant derivative is
\begin{equation}
  (D_1E)^aT^a\coloneqq\partial_1(E^aT^a)
  +\ri g\big[A_1,E^cT^c\big] \,.
  \label{eq:app:adjoint-derivative}
\end{equation}
We assume that the vacuum at $\muQ=0$ is invariant under translations, charge conjugation and parity.

Let $\langle\cdots\rangle_{\muQ}$ denote equal-time expectation values in the physical ground state of the Hamiltonian $H_{\muQ}$.  With $\mathrm P$ denoting path ordering, define the straight color Wilson line and, for $x\ne0$, the correlators
\begin{align}
  U(y,z)&\coloneqq\mathrm P\exp\left[-\ri g\int_z^y dw\,A_1(w)\right] \,,
  \label{eq:app:sunWilson}\\
  G_s(x;\muQ)&\coloneqq\frac{1}{\Nc}
  \big\langle\psi_s^\dagger(x)U(x,0)\psi_s(0)\big\rangle_{\muQ} \,,
  \label{eq:app:suncorrelator}\\
  W(x)&\coloneqq\sum_{s=\pm1}\big[G_s(x;\muQ)-G_s(x;0)\big] \,.
  \label{eq:app:sunobservable}
\end{align}

To derive the factorization of $W(x)$, we first remove the chemical potential from the Dirac operator using the spatial chiral rotation of Ref.~\cite{Kojo:2011en},
\begin{equation}
  \psi(t,x)=e^{\ri\muQ x\gamma^5}\chi(t,x) \,,
  \qquad
  \bar\psi(t,x)=\bar\chi(t,x)e^{\ri\muQ x\gamma^5} \,.
  \label{eq:app:chiralrotation}
\end{equation}
The identity $\gamma^1\gamma^5=\gamma^0$ gives
\begin{equation}
  \bar\psi\big(\ri\gamma^\nu D_\nu+\muQ\gamma^0\big)\psi
  =\bar\chi\,\ri\gamma^\nu D_\nu\chi \,.
  \label{eq:app:rotatedaction}
\end{equation}
The gauge field and the canonical anticommutation relations are unchanged.

We compute the Hamiltonian after the rotation to show that the $\chi$ correlator in $G_s(x;\muQ)$ can be evaluated in the vacuum at $\muQ=0$.  To preserve the antiperiodic boundary conditions for the fermions, we choose $\muQ L=2\pi n$, where $n$ is a positive integer, and take the limit $L\to \infty~(n\to \infty)$ at the end.  Equation~\eqref{eq:app:chiralrotation} then gives $\chi(t,x+L)=-\chi(t,x)$, the same boundary condition as for $\psi$.

We regulate the fermion products at equal time with spatial separation $\epsilon$.  Substituting the free fermion mode expansion gives
\begin{equation}
\begin{aligned}
  \frac{1}{\Nc}\big\langle
  \psi_s^\dagger(x+\epsilon/2)\psi_s(x-\epsilon/2)
  \big\rangle_{\mathrm{free}}
  &=\lim_{\eta\to0}\int_{sp<0}\frac{dp}{2\pi}
  e^{-\ri p\epsilon-\eta|p|}\\
  &=\lim_{\eta\to0}\frac{1}{2\pi(\eta-\ri s\epsilon)}
  =\frac{\ri s}{2\pi\epsilon} \,,
\end{aligned}
  \label{eq:app:reference-contraction}
\end{equation}
where $\eta>0$ is a convergence regulator.  We assume that the interacting correlator with the Wilson line has the same leading pole in the short-distance limit.
Then, let us define the following point-split bilinear,
\begin{equation}
  \mathcal B_s[\psi;\epsilon]
  \coloneqq\psi_s^\dagger(x+\epsilon/2)
  U(x+\epsilon/2,x-\epsilon/2)\psi_s(x-\epsilon/2)
  -\frac{\ri s\Nc}{2\pi\epsilon} \,.
  \label{eq:app:point-split-bilinear}
\end{equation}
Equation~\eqref{eq:app:chiralrotation} gives
\begin{equation}
  \mathcal B_s[\psi;\epsilon]
  =e^{-\ri s\muQ\epsilon}\mathcal B_s[\chi;\epsilon]
  +\frac{\ri s\Nc}{2\pi\epsilon}
  \big(e^{-\ri s\muQ\epsilon}-1\big) \,.
  \label{eq:app:rotated-bilinear}
\end{equation}
Substituting Eq.~\eqref{eq:app:rotated-bilinear} into Eq.~\eqref{eq:app:canonicalhamiltonian}, with a common vacuum energy subtraction for $\psi$ and $\chi$, gives
\begin{equation}
\begin{aligned}
  H_{\muQ}[\psi,A]-H_{\muQ=0}[\chi,A]
  &=\lim_{\epsilon\to0}\int_0^L dx\sum_{s=\pm1}\Big[
  \big(\ri s\partial_\epsilon-\muQ\big)\mathcal B_s[\psi;\epsilon]
  -\ri s\partial_\epsilon\mathcal B_s[\chi;\epsilon]\Big]\\
  &=-\frac{\Nc\muQ^2L}{2\pi} \,.
\end{aligned}
  \label{eq:app:rotatedhamiltonian}
\end{equation}
The Gauss constraint is unchanged.  Because the Hamiltonians differ only by a constant, they have the same ground state.  Taking $n\to\infty$ at fixed $\muQ$ and $x$, Eqs.~\eqref{eq:app:chiralrotation} and \eqref{eq:app:suncorrelator} give
\begin{equation}
  G_s(x;\muQ)
  =e^{-\ri s\muQ x}G_s(x;0) , \qquad x\ne0 \,.
  \label{eq:app:chiralcorrelator}
\end{equation}

We next combine the two chiralities to express $W(x)$ in terms of the vacuum correlator $G_+(x;0)$.  In our gamma convention, charge conjugation acts as $\psi\mapsto-\gamma^5\psi^*$ and $U\mapsto U^*$, while parity acts as $\psi(t,x)\mapsto\gamma^0\psi(t,-x)$ and $U(y,z)\mapsto U(-y,-z)$.  The assumed vacuum symmetries and the equal-time anticommutation relation at $x\ne0$ give
\begin{equation}
  G_s(x;0)=-G_s(-x;0)=-G_{-s}(x;0) .
  \label{eq:app:vacuumchirality}
\end{equation}
Inserting Eqs.~\eqref{eq:app:chiralcorrelator} and \eqref{eq:app:vacuumchirality} into Eq.~\eqref{eq:app:sunobservable} gives
\begin{equation}
  W(x)=-2\ri\sin(\muQ x)G_+(x;0)
  =\frac{\sin(\muQ x)}{\pi x}\mathcal D_c(x) \,,
  \label{eq:app:sunfact}
\end{equation}
where we define
\begin{equation}
  \mathcal D_c(x)\coloneqq-2\pi\ri x\,G_+(x;0) \,,
  \qquad x\ne0 .
  \label{eq:app:vacuumdressing}
\end{equation}
Hermiticity gives $G_s(-x;0)=G_s(x;0)^*$, making $\mathcal D_c$ real and even.  $G_+(x;0)=\ri/(2\pi x)+o(1/x)$ follows from Eq.~\eqref{eq:app:reference-contraction}.  Hence, $\mathcal D_c(x)=1+o(1)$, and its continuous extension has $\mathcal D_c(0)=1$.  Consequently, $W(0)=\muQ/\pi=n_{\mathrm{B}}$ for one flavor.

Fourier transformation of Eq.~\eqref{eq:app:sunfact} gives the convolution in Eq.~\eqref{eq:main:conv}, with Fourier transforms understood as distributions,
\begin{equation}
\begin{aligned}
  f_{\mathrm Q}(q)
  =\int dx\,e^{-\ri qx}W(x)=\int\frac{dq'}{2\pi}\,n_{\mathrm F}(q')\Kone(q-q') \,.
\end{aligned}
  \label{eq:app:sunconvolution}
\end{equation}
Here $n_{\mathrm F}(q')=\Theta(\muQ-|q'|)$ is the free Fermi distribution.  The kernel is independent of $\muQ$ because $\mathcal D_c$ is fixed by the vacuum correlator,
\begin{equation}
  \Kone(q)\coloneqq\int dx\,e^{-\ri qx}\mathcal D_c(x) \,,
  \quad
  \int\frac{dq}{2\pi}\,\Kone(q)=\mathcal D_c(0)=1 \,.
  \label{eq:app:sunkernel}
\end{equation}
No large-$\Nc$ approximation or explicit solution for the vacuum function $\mathcal D_c$ enters the proof.  The kernel shape remains undetermined.  At nonzero quark mass $m_q$, the rotation leaves the term $-m_q\bar\chi e^{2\ri\muQ x\gamma^5}\chi$ in the action.  The argument therefore establishes the factorization for massless continuum quarks, not for nonzero mass or finite lattice spacing.

\bibliography{qknc_new}
\end{document}